\documentclass[11pt,a4paper]{article}
\usepackage{bbm}
\pdfoutput=1
\usepackage{jheppub}
\usepackage{amsmath}
\usepackage{epsfig}
\usepackage{amssymb}
\usepackage{graphics}
\usepackage[active]{srcltx}
\usepackage{epstopdf}
\usepackage{subfigure}
\usepackage{shuffle}
\usepackage[utf8]{inputenc}

\newcommand{\bea}{\begin{eqnarray}}
\newcommand{\eea}{\end{eqnarray}}
\newcommand{\bean}{\begin{eqnarray*}}
\newcommand{\eean}{\end{eqnarray*}}
\newcommand{\nn}{\nonumber \\}

\def\W #1{\widetilde{#1}}

\def\eref#1{(\ref{#1})}

\def\Label#1{\label{#1}%
  \smash{\hbox to0pt{\raise1ex\hbox{\tiny[#1]}\hss}}}

\title{Unified web for expansions of amplitudes}
\author[a]{Kang Zhou}
\affiliation[a]{Center for Gravitation and Cosmology, College of Physical Science and Technology, Yangzhou University,
No. 180, Siwangting Road, Yangzhou, 225009, P.R. China}

\emailAdd{zhoukang@yzu.edu.cn}

\abstract{In this paper, we demonstrate that using differential operators one can construct the complete unified web for expansions of amplitudes for a wide range of theories. We first re-derive the expansion of multi-trace Einstein-Yang-Mills amplitudes to Kleiss-Kuijf basis of color-ordered Yang-Mills amplitudes, by applying proper differential operators which modify the coefficients in the recursive expansion of single-trace Einstein-Yang-Mills amplitudes. Next, through differential operators which act on amplitudes only, we obtain expansions of amplitudes of Yang-Mills theory, Yang-Mills-scalar theory, $\phi^4$ theory, non-linear sigma model, bi-adjoint scalar theory, Born-Infeld theory, Dirac-Born-Infeld theory and special Galileon theory. Then, together with other results in literatures, the complete unified web is achieved. This web for expansions is the dual version of the unified web for differential operators. Thus, connections among amplitudes of a variety of theories, which are reflected by Cachazo-He-Yuan integrands and differential operators previously, can also be represented by expansions. We also find that amplitudes of all theories in the web can be expanded to double color-ordered bi-adjoint scalar amplitudes in the universal double copy formula.}


\begin{document}

\maketitle \flushbottom

\section{Introduction}
\label{introduction}

The modern researches on S-matrix have exposed marvelous properties of scattering amplitudes which are long hidden upon inspecting traditional Feynman rules. Among these properties, one remarkable structure is the deep connections between amplitudes of different theories. These unexpected connections were first indicated by the well known Kawai-Lewellen-Tye (KLT) relation \cite{Kawai:1985xq} and Bern-Carrasco-Johansson (BCJ) color-kinematic duality \cite{Bern:2008qj} which factorize tree-level amplitudes of gravity (GR) as the double copy of tree-level amplitudes of Yang-Mills theory (YM). The Cachazo-He-Yuan (CHY) formula proposed in 2013 links a wider range of theories together \cite{Cachazo:2013gna,Cachazo:2013hca, Cachazo:2013iea, Cachazo:2014nsa,Cachazo:2014xea}. Tree-level amplitudes in the CHY formula are expressed as multi-dimensional contour
integral over auxiliary variables as
\bea
{\cal A}_n=\int d\mu_n\,{\cal I}^{\rm CHY}\,.~~~~\label{CHY-0}
\eea
In this formula, the measure part $d\mu_n$ is universal for all theories, while the integrand ${\cal I}^{\rm CHY}$ depends on the theory under consideration. As exhibited in \cite{Cachazo:2014xea}, through operations so called dimension reduction, squeezing, and generalized dimension reduction, CHY integrands for a variety of theories including Einstein-Yang-Mills (EYM), Einstein-Maxwell (EM), Born-Infeld (BI), YM, Yang-Mills-scalar (YMS), $\phi^4$, non-linear sigma model (NLSM), bi-adjoint scalar (BAS), Dirac-Born-Infeld (DBI), as well as special Galileon (SG), can be generated from integrands for GR. Recently, the same web for connections has been reproduced by introducing differential operators \cite{Cheung:2017ems}. In this framework, the tree-level GR amplitudes are transmuted to tree-level amplitudes of other theories via proper differential operators which act on kinematic variables. The connection between two pathes which lead to the same web picture is then exposed in \cite{Zhou:2018wvn,Bollmann:2018edb}.

Another significant reflection of these connections is that tree-level amplitudes of one theory can be expanded to thoes of another theory, which have been studied in various literatures recently, especially for the expansion of tree-level EYM amplitudes to tree-level YM ones \cite{Stieberger:2016lng,Schlotterer:2016cxa,Chiodaroli:2017ngp,DelDuca:1999rs,Nandan:2016pya,delaCruz:2016gnm,Fu:2017uzt,Teng:2017tbo,Du:2017kpo,
Du:2017gnh}. To obtain the coefficients in expansions, several efforts have been devoted in above literatures. Among these methods, the recent proposed approach, which based on differential operators \cite{Feng:2019tvb}, indicates that the unifying relations of amplitudes described by differential operators in \cite{Cheung:2017ems,Zhou:2018wvn,Bollmann:2018edb} and the expansions of amplitudes relate to each other. As discussed in \cite{Feng:2019tvb}, expansions of single-trace EYM and GR amplitudes to color-ordered YM amplitudes can be reached by solving differential equations indicated by differential operators, together with considering the gauge invariance of gravitons. Then, by applying appropriate differential operators on above two expansions, expansions of amplitudes of BI, EM, $\phi^4$, sYMS (special Yang-Mills-scalar) and DBI can be derived straightforwardly \cite{Feng:2019tvb,Hu:2019qdq}. In other words, expansions listed above emergence naturally from the unified web for differential operators.

\begin{figure}
  \centering
  \includegraphics[width=12cm]{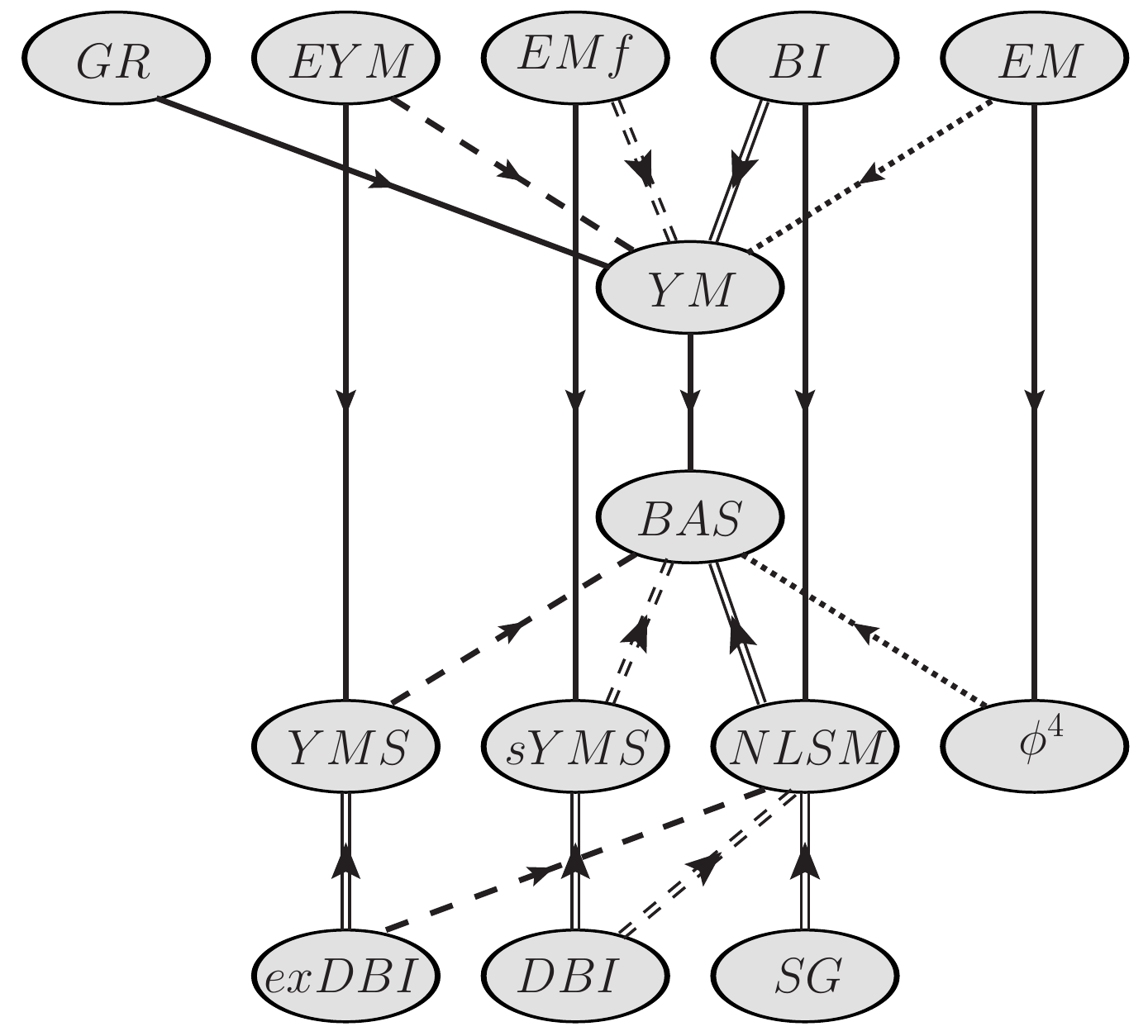} \\
  \caption{Unified web for expansions}\label{web}
\end{figure}

Inspired by connections between the unified web for differential operators and the expansions of amplitudes, it is natural to expect that a new unified web for expansions which includes all theories in the previous web for differential operators, can be established. In this paper, such a new web is constructed by applying only differential operators, i.e, only information in the previous web. The procedure of the construction is, (1) derive expansions of GR and single-trace EYM amplitudes to YM amplitudes in the ordered splitting formula by solving differential equations, (2) derive expansions of BI, EM and multi-trace EYM amplitudes to YM amplitudes in the ordered splitting formula by applying proper differential operators on expansions obtained in the step (1), (3) extract coefficients of YM amplitudes from expansions in the ordered splitting formula obtained in the step (2), (4) apply proper operators on expansions obtained in the step (3) to get expansions for other theories. Some manipulations in above four steps are already finished in literatures, and the remaining part are completed in the current paper. Through out the procedure above, the only backgrounds are relations between amplitudes, which are represented by differential operators. No other prior assumption is required. The obtained expansions can be organized into the unified web as shown in Fig. \ref{web}.
In this web, directions of arrows point to basis in the expansion, and different kinds of lines correspond to different kinds of coefficients.
This web describes connections between amplitudes of different theories by expansions, and serves as the dual version of the unified web for differential operators: two webs include same theories, and there are one to one mappings between differential operators and coefficients in expansions. In the dual version of unifying relations described by expansions, amplitudes of all other theories arise as the linear combinations of BAS amplitudes whose external particles carry lowest spin among theories in the web. Indeed, expansions in the web can be unified into the double copy formula
\bea
{\cal A}=\sum_{\sigma}\sum_{\sigma'}\,{\cal C}(\sigma){\cal A}_{\rm BAS}(1,\sigma,n;1,\sigma',n){\cal C}(\sigma')\,,
\eea
where the double color-ordered BAS amplitudes ${\cal A}_{\rm BAS}(1,\sigma,n;1,\sigma',n)$ form the propagator matrix. Numerators ${\cal C}(\sigma)$ and ${\cal C}(\sigma')$ are created by applying proper differential operators on BCJ numerators of YM amplitudes.

This paper is organized as follows. In section \ref{back}, we give a brief review of necessary backgrounds, including differential operators and unifying relations for amplitudes, the choices of basis in the expansions, as well as the recursive expansions for GR and single-trace EYM amplitudes, which are crucial for later discussions. We also point out that the basis and the recursive expansions can be obtained only through the knowledge of differential operators. In section \ref{multi-trace EYM}, we give the derivation of the recursive expansions of multi-trace EYM amplitudes by applying differential operators, to finish the step (2) in the procedure of constructing the unified web. Section \ref{coefficients} devotes to systematic algorithms of evaluating coefficients of KK basis, and relations among coefficients for different theories. In section \ref{unified web}, we establish the whole web for expansions, and organize these expansions into the double copy formula. Finally, we end with a summary and discussion in section \ref{summary}.

\section{Some backgrounds}
\label{back}

For reader's convenience, we review some backgrounds which are crucial for discussions in next sections. In the first subsection, we introduce differential operators proposed by Cheung, Shen and Wen, as well as unifying relations among a variety of amplitudes which are described by these operators. In the second subsection, we discuss the choices of basis for expansions, and emphasize that basis can be determined only through differential operators. In the third subsection, we rapidly review the recursive expansions for single-trace EYM and GR amplitudes, which emergence naturally from connections between amplitudes described by differential operators. Some notations which are used through out the paper will also be provided in this section.

\subsection{Differential operators}

The differential operators introduced by cheung, Shen and Wen transmute tree-level amplitudes of one theory to amplitudes of other theories \cite{Cheung:2017ems,Zhou:2018wvn,Bollmann:2018edb}. Three kinds of basic operators are defined as follows:
\begin{itemize}
\item (1) Trace operator:
\bea
{\cal T}^\epsilon[ij]\equiv {\partial\over\partial(\epsilon_i\cdot\epsilon_j)}\,,
\eea
where $\epsilon_i$ is the polarization vector of $i$th external leg. The up index $\epsilon$ means the operators are defined through polarization vectors $\epsilon_i$.
\item (2) Insertion operator:
\bea
{\cal T}^\epsilon_{ikj}\equiv {\partial\over\partial(\epsilon_k\cdot k_i)}-{\partial\over\partial(\epsilon_k\cdot k_j)}\,,~~~~\label{defin-insertion}
\eea
where $k_i$ denotes the momentum of the $i$th external leg. When applying to physical amplitudes, the insertion operator ${\cal T}^\epsilon_{ik(i+1)}$ inserts the external leg $k$ between $i$th and $(i+1)$th external legs in the color-ordering $(\cdots,i,i+1,\cdots)$. For general ${\cal T}^\epsilon_{ikj}$ with $i<j$, one can use the definition \eref{defin-insertion} to dived ${\cal T}^\epsilon_{ikj}$ as
\bea
{\cal T}^\epsilon_{ikj}={\cal T}^\epsilon_{ik(i+1)}+{\cal T}^\epsilon_{(i+1)k(i+2)}+\cdots+{\cal T}^\epsilon_{(j-1)kj}\,.
\eea
In the above expression, each ${\cal T}^\epsilon_{ak(a+1)}$ on the RHS can be interpreted as inserting the leg $k$ between $a$ and $(a+1)$.
Consequently, the effect of ${\cal T}^\epsilon_{ikj}$ can be understood as inserting $k$ between $i$ and $j$ in the color-ordering $(\cdots,i,\cdots,j,\cdots)$, and summing over all possible positions together.
\item (3) Longitudinal operator:
\bea
{\cal L}^\epsilon_i\equiv \sum_{j\neq i}\,k_i\cdot k_j{\partial\over\partial(\epsilon_i\cdot k_j)}\,,~~~~~~~~
{\cal L}^\epsilon_{ij}\equiv -k_i\cdot k_j{\partial\over\partial(\epsilon_i\cdot \epsilon_j)}\,.
\eea
\end{itemize}

By using products of these three kinds of basic operators, one can transmute amplitudes of one theory into those of other theories. Three combinatory operators which are products of basic operators are defined as follows:
\begin{itemize}
\item (1) For a length-$m$ ordered set $\pmb{\bar{\alpha}}=\{\alpha_1,\alpha_2,\cdots,\alpha_m\}$ of external particles, the operator ${\cal T}^\epsilon[\pmb{\bar{\alpha}}]$
is given as
\bea
{\cal T}^\epsilon[\pmb{\bar{\alpha}}]\equiv {\cal T}^\epsilon[\alpha_1\alpha_m]\cdot\prod_{i=2}^{m-1}\,{\cal T}^\epsilon_{\alpha_{i-1}\alpha_i\alpha_m}\,.
\eea
It fixes $\alpha_1$ and $\alpha_m$ at two ends of the color-ordering via the operator ${\cal T}^\epsilon[\alpha_1\alpha_m]$, and inserts other elements between them by insertion operators.
The operator ${\cal T}^\epsilon[\pmb{\bar{\alpha}}]$ is also called the trace operator since it generates the color-ordering $(\alpha_1,\alpha_2,\cdots,\alpha_m)$. The interpretation of insertions operators indicates that ${\cal T}^\epsilon[\pmb{\bar{\alpha}}]$ has various equivalent formulae, for example
\bea
& &{\cal T}^\epsilon[\pmb{\bar{\alpha}}]\equiv {\cal T}^\epsilon[\alpha_1\alpha_m]\cdot\prod_{i=m-1}^{2}\,{\cal T}^\epsilon_{\alpha_{1}\alpha_i\alpha_{i+1}}\,,\nn
& &{\cal T}^\epsilon[\pmb{\bar{\alpha}}]\equiv {\cal T}^\epsilon[\alpha_1\alpha_m]\cdot {\cal T}^\epsilon_{\alpha_{1}\alpha_2\alpha_m}\cdot{\cal T}^\epsilon_{\alpha_{2}\alpha_{m-1}\alpha_m}\cdot\prod_{i=3}^{m-2}\,{\cal T}^\epsilon_{\alpha_{i-1}\alpha_i\alpha_{m-1}}\,,
\eea
and so on.
\item (2) For $n$-point amplitudes, the operator ${\cal L}^\epsilon$
is defined as
\bea
{\cal L}^\epsilon\equiv\prod_i\,{\cal L}^\epsilon_i,~~~~~~~~\W{\cal L}^\epsilon\equiv\sum_{\rho\in{\rm pair}}\,\prod_{i,j\in\rho}\,{\cal L}^\epsilon_{ij}\,.
\eea
Two definitions ${\cal L}^\epsilon$ and $\W{\cal L}^\epsilon$ are not equivalent to each other at the algebraic level. However, when acting on proper on-shell
physical amplitudes, two combinations
${\cal L}^\epsilon\cdot{\cal T}^\epsilon[ab]$ and $\W{\cal L}^\epsilon\cdot{\cal T}^\epsilon[ab]$, with subscripts of ${\cal L}^\epsilon_i$ and ${\cal L}^\epsilon_{ij}$
run through all nodes in $\{1,2,\cdots,n\}\setminus\{a,b\}$, give the same effect which can be interpreted physically.
\item (3) For a length-$2m$ set $\pmb{I}$, the operator ${\cal T}^\epsilon_{{\cal X}_{2m}}$ is defined as
\bea
{\cal T}^\epsilon_{{\cal X}_{2m}}\equiv\sum_{\rho\in{\rm pair}}\,\prod_{i_k,j_k\in\rho}\,\delta_{I_{i_k}I_{j_k}}{\cal T}^\epsilon_{i_kj_k}\,,
\eea
where $\delta_{I_{i_k}I_{j_k}}$ forbids the interaction between particles with different flavors. For the special case $2m$ particles do not carry any flavor, the operator ${\cal T}^\epsilon_{X_{2m}}$ is defined by removing $\delta_{I_{i_k}I_{j_k}}$,
\bea
{\cal T}^\epsilon_{X_{2m}}\equiv\sum_{\rho\in{\rm pair}}\,\prod_{i_k,j_k\in\rho}\,{\cal T}^\epsilon_{i_kj_k}\,.
\eea
\end{itemize}
The explanation for the notation $\sum_{\rho\in{\rm pair}}\,\prod_{i_k,j_k\in\rho}$ is in order. Let $\Gamma$ be the set of all partitions of the set $\{1,2,\cdots, 2m\}$ into pairs without regard to the order.
An element in $\Gamma$ can be written as
\bea
\rho=\{(i_1,j_1),(i_2,j_2),\cdots,(i_m,j_m)\}\,,
\eea
with conditions $i_i<i_2<\cdots<i_m$ and $i_t<j_t,\,\forall t$. Then, $\prod_{i_k,j_k\in\rho}$ stands for the product of ${\cal T}^\epsilon_{i_kj_k}$
for all pairs $(i_k,j_k)$ in $\rho$, and $\sum_{\rho\in{\rm pair}}$ denotes the summation over all partitions.

The combinatory operators exhibited above unifies tree-level amplitudes of a wide range of theories together, by translating the GR amplitudes into amplitudes of other theories, formally expressed as
\bea
{\cal A}={\cal O}^\epsilon{\cal O}^{\W\epsilon}{\cal A}^{\epsilon,\W\epsilon}_{\rm GR}\,.~~~~\label{fund-uni-diff}
\eea
Operators ${\cal O}^\epsilon$ and ${\cal O}^{\W\epsilon}$ for different theories are listed in Table \ref{tab:unifying}.
\begin{table}[!h]
\begin{center}
\begin{tabular}{c|c|c}
Amplitude& ${\cal O}^\epsilon$  & ${\cal O}^{\W\epsilon}$ \\
\hline
${\cal A}_{{\rm GR}}^{\epsilon,\W\epsilon}(\pmb{H}_n)$ & $\mathbb{I}$ & $\mathbb{I}$  \\
${\cal A}_{{\rm EYM}}^{\epsilon,\W\epsilon}(\pmb{\bar{1}}|\cdots|\pmb{\bar{r}}||\pmb{H}_{n-(|\pmb{1}|+\cdots+|\pmb{r}|)})$ & $\mathbb{I}$ & ${\cal T}^{\W\epsilon}[\pmb{\bar{1}}]\cdots{\cal T}^{\W\epsilon}[\pmb{\bar{r}}]$  \\
${\cal A}_{{\rm EMf}}^{\epsilon,\W\epsilon}(\pmb{P}_{2m}||\pmb{H}_{n-2m})$ & $\mathbb{I}$ & ${\cal T}^{\W\epsilon}_{{\cal X}_{2m}}$  \\
${\cal A}_{{\rm EM}}^{\epsilon,\W\epsilon}(\pmb{P}_{2m}||\pmb{H}_{n-2m})$ & $\mathbb{I}$ & ${\cal T}^{\W\epsilon}_{X_{2m}}$  \\
${\cal A}_{{\rm BI}}^\epsilon(\pmb{P}_n)$ & $\mathbb{I}$ & ${\cal L}^{\W\epsilon}\cdot{\cal T}^{\W\epsilon}[ab]$ \\
${\cal A}_{{\rm YM}}^\epsilon(i_1,\cdots,i_n)$ & $\mathbb{I}$ & ${\cal T}^{\W\epsilon}[i_1\cdots i_n]$  \\
${\cal A}_{{\rm YMS}}^{\W\epsilon}(\pmb{\bar{1}}|\cdots|\pmb{\bar{r}}||\pmb{G}_{n-(|\pmb{1}|+\cdots+|\pmb{r}|)};i'_1,\cdots,i'_n)$ & ${\cal T}^{\epsilon}[i'_1\cdots i'_n]$ & ${\cal T}^{\W\epsilon}[\pmb{\bar{1}}]\cdots{\cal T}^{\W\epsilon}[\pmb{\bar{r}}]$ \\
${\cal A}_{{\rm sYMS}}^{\W\epsilon}(\pmb{S}_{2m}||\pmb{G}_{n-2m};i'_1,\cdots,i'_n)$ &  ${\cal T}^{\epsilon}[i'_1\cdots i'_n]$ & ${\cal T}^{\W\epsilon}_{{\cal X}_{2m}}$ \\
${\cal A}_{\phi^4}(i'_1,\cdots,i'_n)$ &  ${\cal T}^{\epsilon}[i'_1\cdots i'_n]$ & ${\cal T}^{\W\epsilon}_{X_n}$ \\
${\cal A}_{{\rm NLSM}}(i'_1,\cdots,i'_n)$ & ${\cal T}^{\epsilon}[i'_1\cdots i'_n]$ & ${\cal L}^{\W\epsilon}\cdot{\cal T}^{\W\epsilon}[ab]$ \\
${\cal A}_{{\rm BS}}(i_1,\cdots,i_n;i'_1,\cdots,i'_n)$ &  ${\cal T}^{\epsilon}[i'_1\cdots i'_n]$ & ${\cal T}^{\W\epsilon}[i_1\cdots i_n]$ \\
${\cal A}_{{\rm exDBI}}^{\W\epsilon}(\pmb{\bar{1}}|\cdots|\pmb{\bar{r}}||\pmb{P}_{n-(|\pmb{1}|+\cdots+|\pmb{r}|)})$ &  ${\cal L}^{\epsilon}\cdot{\cal T}^{\epsilon}[a'b']$ & ${\cal T}^{\W\epsilon}[\pmb{\bar{1}}]\cdots{\cal T}^{\W\epsilon}[\pmb{\bar{r}}]$ \\
${\cal A}_{{\rm DBI}}^{\W\epsilon}(\pmb{S}_{2m}||\pmb{P}_{n-2m})$ &${\cal L}^{\epsilon}\cdot{\cal T}^{\epsilon}[a'b']$ & ${\cal T}^{\W\epsilon}_{{\cal X}_{2m}}$ \\
${\cal A}_{{\rm SG}}(\pmb{S}_n)$ &  ${\cal L}^{\epsilon}\cdot{\cal T}^{\epsilon}[a'b']$ & ${\cal L}^{\W\epsilon}\cdot{\cal T}^{\W\epsilon}[ab]$ \\
\end{tabular}
\end{center}
\caption{\label{tab:unifying}Unifying relations for differential operators}
\end{table}
In this table, all amplitudes include $n$ external particles. EMf denotes the EM theory that photons carry flavors, sYMS stands for the special YMS theory, and exDBI denotes the extended DBI theory. The symbol $\mathbb{I}$ stands for the identical operator. Notations $\pmb{H}_a$, $\pmb{P}_a$, $\pmb{G}_a$ and $\pmb{S}_a$
denote sets of gravitons, photons, gluons and scalars respectively, where the subscript denotes the length of the set. Sometimes we will omit the subscribe of a set
if the length of this set is not important. Through this paper, a bold number or letter stands for a set, and $\pmb{\bar{}}$ denotes that the set is ordered. The notation $|\pmb{\alpha}|$ stands for the length of the set $\pmb{\alpha}$. We use ${\cal A}_{{\rm YMS}}^{\W\epsilon}(\pmb{\bar{1}}|\cdots|\pmb{\bar{r}}||\pmb{G}_{n-(|\pmb{1}|+\cdots+|\pmb{r}|)};i'_1,\cdots,i'_n)$ as the example to explain notations $;\cdots$, $|$ and $||$.
The notation $;\cdots$ denotes the additional color-ordering among all external particles, such as $(i'_1,\cdots,i'_n)$ among all scalars and gluons in the example. Notations $|$ and $||$ are used to separate different sets of external particles. Sets on two sides of $|$ belong to the same kind of particles, such as sets $\pmb{\bar{i}}$ of scalars for different traces. Sets on two sides of $||$ belong to different kinds of particles with the convention: particles on the LHS of $||$ carry lower spin. In our example, the LHS of $||$ are sets of scalars while the RHS is the set of gluons. The up index of
${\cal A}$ denotes the polarization vectors of external particles. In the cases amplitudes include external gravitons, the rule is: the previous polarization vectors are carried by all particles,
while the later ones are only carried by gravitons. For instance, in the notation ${\cal A}_{{\rm EMf}}^{\epsilon,\W\epsilon}(\pmb{P}_{2m}||\pmb{H}_{n-2m})$, $\epsilon_i$ are carried by both photons and gravitons, while $\W\epsilon_i$ are only carried by gravitons.

In Table \ref{tab:unifying}, two copies labeled by polarization vectors $\epsilon$ and $\W\epsilon$ are exchangeable. As an example, YM amplitudes carry the polarization vectors $\W\epsilon$ can be generated by
\bea
{\cal A}^{\W\epsilon}_{\rm YM}(i_1,\cdots,i_n)={\cal T}^\epsilon[i_1,\cdots,i_n]{\cal A}^{\epsilon,\W\epsilon}_{\rm GR}(\pmb{H}_n)\,.
\eea
All relations between amplitudes of different theories can be extracted from Table \ref{tab:unifying}. For example, from relations
\bea
& &{\cal A}^\epsilon_{\rm BI}(\pmb{P}_n)={\cal L}^{\W\epsilon}\cdot{\cal T}^{\W\epsilon}[ab]{\cal A}^{\epsilon,\W\epsilon}_{\rm GR}(\pmb{H}_n)\,,\nn
& &{\cal A}_{\rm NLSM}(i'_1,\cdots,i'_n)={\cal T}^\epsilon[i'_1,\cdots,i'_n]\Big({\cal L}^{\W\epsilon}\cdot{\cal T}^{\W\epsilon}[ab]\Big){\cal A}^{\epsilon,\W\epsilon}_{\rm GR}(\pmb{H}_n)\,,
\eea
one can get
\bea
& &{\cal A}_{\rm NLSM}(i'_1,\cdots,i'_n)={\cal T}^\epsilon[i'_1,\cdots,i'_n]{\cal A}^\epsilon_{\rm BI}(\pmb{P}_n)\,.
\eea
Thus, the formula \eref{fund-uni-diff} is the underlying foundation of the unified web for differential operators.

\subsection{KK basis and the generalization}
\label{determin-basis}

When considering the expansion of amplitudes, an essential condition for basis is the completeness. In this subsection, we will show that basis satisfying this requirement can be determined only through differential operators.

Let us begin with introducing the KK basis constituted by color-ordered YM amplitudes, which is related to the well known
Kleiss-Kuijf (KK) relation \cite{Kleiss:1988ne}
\bea
{\cal A}^{\epsilon}_{\rm YM}(1,\pmb{\bar{a}},n,\pmb{\bar{b}})=\sum_{\shuffle}\,(-)^{|\pmb{b}|}{\cal A}^{\epsilon}_{\rm YM}(1,\pmb{\bar{a}}\shuffle\pmb{\bar{b}}^T,n)\,,~~~~\label{KK}
\eea
where $\pmb{\bar{a}}$ and $\pmb{\bar{b}}$ are two ordered subsets of external gluons,
and $\pmb{\bar{b}}^T$ denotes the set obtained from $\pmb{\bar{b}}$ by reversing the original ordering. The summation is over all
possible shuffles of two ordered subsets, i.e., all permutations in the set $\pmb{\bar{a}}\cup\pmb{\bar{b}}^T$ while preserving the ordering
of $\pmb{\bar{a}}$ and $\pmb{\bar{b}}^T$.
The KK relation \eref{KK} indicates that an arbitrary color-ordered YM amplitude can be expanded into color-ordered YM amplitudes ${\cal A}^\epsilon_{\rm YM}(1,\sigma_2,\cdots,\sigma_{n-1},n)$. Thus, the set of $(n-2)!$ color-ordered YM amplitudes, with two legs are fixed at two ends in the color-orderings, satisfies the condition of completeness.
Thus, the set of amplitudes ${\cal A}^\epsilon_{\rm YM}(1,\sigma_2,\cdots,\sigma_{n-1},n)$ can be chosen as the complete basis, which is called the KK basis.

As discussed in \cite{Feng:2019tvb}, the KK relation can be derived by applying differential operators. Indeed, the KK relation can be regarded as the consequence of the property of the operator ${\cal T}^{\W\epsilon}[\pmb{\bar{\alpha}}]$.
To show this, let us rewrite the operator ${\cal T}^{\W\epsilon}[\pmb{\bar{\alpha}}]$ for the length-$n$ set $\pmb{\bar{\alpha}}$ as
\bea
{\cal T}^\epsilon[\pmb{\bar{\alpha}}]&\equiv & {\cal T}^\epsilon[\alpha_1\alpha_n]\cdot\prod_{i=2}^{n-1}\,{\cal T}^\epsilon_{\alpha_{i-1}\alpha_i\alpha_n}\nn
&=&{\cal T}^\epsilon[\alpha_1\alpha_n]\cdot\Big(\prod_{i=2}^{k}\,{\cal T}^\epsilon_{\alpha_{i-1}\alpha_i\alpha_n}\Big)\cdot\Big(\prod_{j=k+1}^{n-1}\,{\cal T}^\epsilon_{\alpha_{j-1}\alpha_j\alpha_n}\Big)\nn
&=&{\cal T}^\epsilon[\alpha_1\alpha_n]\cdot\Big(\prod_{i=2}^{k}\,{\cal T}^\epsilon_{\alpha_{i-1}\alpha_i\alpha_n}\Big)\cdot\Big((-)^{n-k-1}\prod_{j=k+1}^{n-1}\,{\cal T}^\epsilon_{\alpha_n\alpha_j\alpha_{j-1}}\Big)\,.
~~~~\label{pro-trace}
\eea
The operator ${\cal T}^\epsilon[\alpha_1\alpha_n]\cdot\Big(\prod_{i=2}^{k}\,{\cal T}^\epsilon_{\alpha_{i-1}\alpha_i\alpha_n}\Big)$
creates the color-ordering $(\alpha_1,\alpha_2,\cdots,\alpha_k,\alpha_n)$, which is equivalent to $(\alpha_n,\alpha_1,\alpha_2,\cdots,\alpha_k)$
by the cyclic symmetry, and the operator $(-)^{n-k-1}\prod_{j=k+1}^{n-1}\,{\cal T}^\epsilon_{\alpha_n\alpha_j\alpha_{j-1}}$
is interpreted as inserting $\{\alpha_{n-1},\alpha_{n-2},\cdots,\alpha_{k+1}\}$ between $\alpha_n$ and $\alpha_k$. Choosing $\alpha_n=1$, $\alpha_k=n$,
$\pmb{\bar{a}}=\{\alpha_1,\cdots,\alpha_{k-1}\}$, $\pmb{\bar{b}}=\{\alpha_{k+1},\cdots,\alpha_{n-1}\}$, and applying this operator on the $n$-point GR
amplitude ${\cal A}^{\epsilon,\W\epsilon}_{\rm GR}(\pmb{H}_n)$, one get the KK relation \eref{KK} immediately.

The algebraic relation \eref{pro-trace} is general, thus one can expect that the generalized KK relations exist among color-ordered amplitudes
of other theories beyond YM.
Replacing ${\cal A}^{\epsilon,\W\epsilon}_{\rm GR}(\pmb{H}_n)$ in the above derivation by ${\cal A}^{\epsilon,\W\epsilon}_{\rm EYM}(\pmb{\bar{1}}|\cdots|\pmb{\bar{r}}||\pmb{H}_{n-(|\pmb{1}|+\cdots+|\pmb{r}|)})$, ${\cal A}^{\epsilon,\W\epsilon}_{\rm EMf}(\pmb{P}_{2m}||\pmb{H}_{n-2m})$, ${\cal A}^{\epsilon,\W\epsilon}_{\rm EM}(\pmb{P}_n)$, ${\cal A}^\epsilon_{\rm BI}(\pmb{P}_n)$ and ${\cal A}^\epsilon_{\rm YM}(i'_1,\cdots,i'_n)$ respectively, we get the generalized
KK relations as follows
\bea
& &{\cal A}^{\W\epsilon}_{\rm YMS}(\pmb{\bar{1}}|\cdots|\pmb{\bar{r}}||\pmb{H}_{n-(|\pmb{1}|+\cdots+|\pmb{r}|)};1,\pmb{\bar{a}},n,\pmb{\bar{b}})=\sum_{\shuffle}\,(-)^{|\pmb{b}|}{\cal A}^{\W\epsilon}_{\rm YMS}(\pmb{\bar{1}}|\cdots|\pmb{\bar{r}}||\pmb{G}_{n-(|\pmb{1}|+\cdots+|\pmb{r}|)};1,\pmb{\bar{a}}\shuffle\pmb{\bar{b}}^T,n)\,,\nn
& &{\cal A}^{\W\epsilon}_{\rm sYMS}(\pmb{S}_{2m}||\pmb{G}_{n-2m};1,\pmb{\bar{a}},n,\pmb{\bar{b}})=\sum_{\shuffle}\,(-)^{|\pmb{b}|}{\cal A}^{\W\epsilon}_{\rm sYMS}(\pmb{P}_{2m}||\pmb{G}_{n-2m};1,\pmb{\bar{a}}\shuffle\pmb{\bar{b}}^T,n)\,,\nn
& &{\cal A}_{\phi^4}(1,\pmb{\bar{a}},n,\pmb{\bar{b}})=\sum_{\shuffle}\,(-)^{|\pmb{b}|}{\cal A}_{\phi^4}(1,\pmb{\bar{a}}\shuffle\pmb{\bar{b}}^T,n)\,,\nn
& &{\cal A}_{\rm NLSM}(1,\pmb{\bar{a}},n,\pmb{\bar{b}})=\sum_{\shuffle}\,(-)^{|\pmb{b}|}{\cal A}_{\rm NLSM}(1,\pmb{\bar{a}}\shuffle\pmb{\bar{b}}^T,n)\,,\nn
& &{\cal A}_{\rm BAS}(1,\pmb{\bar{a}},n,\pmb{\bar{b}};i'_1,\cdots,i'_n)=\sum_{\shuffle}\,(-)^{|\pmb{b}|}{\cal A}_{\rm BAS}(1,\pmb{\bar{a}}\shuffle\pmb{\bar{b}}^T,n;i'_1,\cdots,i'_n)\,.
\eea
Notice that the BAS amplitudes carry two color-orderings, thus we have\footnote{In general, the fixed legs at two ends for two color-orderings can be chosen independently, here we choose $1$ and $n$ for both two color-orderings for convenience.}
\bea
{\cal A}_{\rm BAS}(1,\pmb{\bar{a}},n,\pmb{\bar{b}};1,\pmb{\bar{c}},n,\pmb{\bar{d}})=\sum_{\shuffle}\,(-)^{|\pmb{b}|+|\pmb{d}|}{\cal A}_{\rm BAS}(1,\pmb{\bar{a}}\shuffle\pmb{\bar{b}}^T,n;1,\pmb{\bar{c}}\shuffle\pmb{\bar{d}}^T,n)\,.
\eea
Thus, these amplitudes share the completeness of KK basis, due to the generalized KK relations. Consequently, we get the following basis
\bea
& &{\cal A}^{\W\epsilon}_{\rm YMS}(\pmb{\bar{1}}|\cdots|\pmb{\bar{r}}||\pmb{G}_{n-(|\pmb{1}|+\cdots+|\pmb{r}|)};1,\sigma_2,\cdots,\sigma_{n-1},n)\,,~~~~~~ {\cal A}^{\W\epsilon}_{\rm sYMS}(\pmb{S}_{2m}||\pmb{G}_{n-2m};1,\sigma_2,\cdots,\sigma_{n-1},n)\,,\nn
& &{\cal A}_{\phi^4}(1,\sigma_2,\cdots,\sigma_{n-1},n)\,,~~~~~~ {\cal A}_{\rm NLSM}(1,\sigma_2,\cdots,\sigma_{n-1},n)\,,\nn
& &{\cal A}_{\rm BAS}(1,\sigma_2,\cdots,\sigma_{n-1},n;1,\sigma'_2,\cdots,\sigma'_{n-1},n)\,,
\eea
which are the generalization of the original KK basis. Notice that basis in the first line are reducible. For example, when expanding the color-ordered sYMS amplitude ${\cal A}^{\W\epsilon}_{\rm sYMS}(\emptyset||\pmb{G}_n;i_1,\cdots,i_n)$, the corresponding basis are
${\cal A}^{\W\epsilon}_{\rm sYMS}(\emptyset||\pmb{G}_n;1,\sigma_2,\cdots,\sigma_{n-1},n)$, with fixed subsets of external scalars and gluons.

In the above discussion, we have determined basis only through differential operators. Indeed, there is another alternative choice of basis with three external legs are fixed, due to the well-known BCJ relations \cite{Bern:2008qj} \footnote{The BCJ relations can also be generalized to color-ordered amplitudes beyond YM ones, since one can applying differential operators which only modify amplitudes on the original BCJ relations.}. However, the basis discussed in this subsection are more suitable for expansions dual to Table \ref{tab:unifying}, since they emergence naturally from differential operators.

\subsection{Recursive expansions for single-trace EYM amplitudes and GR amplitudes}

In this subsection, we review the recursive expansions for single-trace EYM amplitudes and GR amplitudes. We first emphasize that these two expansions can be obtained naturally from Table \ref{tab:unifying}. More explicitly, as discussed in \cite{Feng:2019tvb}, two expansions can be derived by solving differential equations provided by differential operators, together with considering the gauge invariance. The derivation in \cite{Feng:2019tvb} requires no prior assumption for basis. In their method, coefficients of basic Lorentz invariance building blocks such as $\W\epsilon_i\cdot k_i$ and $\W\epsilon_i\cdot\W\epsilon_j$, which can be solved from differential equations, lead to KK basis automatically. This feature supports our claim that the KK basis is the natural choice for expansions dual to Table \ref{tab:unifying}.

The recursive expansion expresses the single-trace EYM amplitude as the sum of single-trace EYM amplitudes with less gravitons, which is given as
\bea
{\cal A}^{\epsilon,\W\epsilon}_{\rm{EYM}}
(1,2,\cdots,n||\{h_1,\pmb{H}\})&=&\sum_{\pmb{\bar{\alpha}}}\,C(\pmb{\bar{\alpha}}){\cal A}^{\epsilon,\W\epsilon}_{\rm EYM}(1,\{2,\cdots,n-1\}\shuffle\{\pmb{\bar{\alpha}},h_1\},n||\pmb{H}
\setminus\pmb{\alpha})\,,~~~~\label{multi-trace-ori}
\eea
where
\bea
C(\pmb{\bar{\alpha}})=\W\epsilon_{h_1}\cdot \W f_{\alpha_m}\cdot \W f_{\alpha_{m-1}}\cdots \W f_{\alpha_1}\cdot Y_{\alpha_1}\,.~~~~\label{defin-C}
\eea
Here $\pmb{\bar{\alpha}}$ is a length-$m$ ordered set $\pmb{\bar{\alpha}}=\{\alpha_1,\alpha_2,\cdots,\alpha_m\}$,
with $m$ runs through $0$ until $|\pmb{H}|$. When $m=0$, $C(\emptyset)=\W\epsilon_{h_1}\cdot Y_{h_1}$.
When $m=|\pmb{H}|+|\pmb{2}|$, the EYM amplitudes ${\cal A}^{\epsilon,\W\epsilon}_{\rm EYM}(1,\{2,\cdots,n-1\}\shuffle\{\pmb{\bar{\alpha}},h_1\},n||\pmb{H}/\pmb{\alpha})$ are reduced to the pure YM ones ${\cal A}^{\epsilon}_{\rm YM}(1,\{2,\cdots,n-1\}\shuffle\{\pmb{\bar{\alpha}},h_1\},n)$.
The strength tensors are defined as
\bea
f_i^{\mu\nu}\equiv \epsilon^\mu_i k^\nu_i-k^\mu_i\epsilon^\nu_i\,,~~~~\W f_i^{\mu\nu}\equiv \W\epsilon^\mu_i k^\nu_i-k^\mu_i\W\epsilon^\nu_i\,.
\eea
The combinatory momentum $Y_{\alpha_1}$ is defined as the sum of momenta of gluon-legs on the LHS of $\alpha_1$ in the color-ordering.
The expansion of single-trace EYM amplitudes to KK basis in the ordered splitting formula can be obtained by applying the expansion \eref{defin-C} recursively.

The recursive expansion for GR amplitudes is given by \footnote{The explicit formula of the recursive expansion of GR amplitudes have not been provided in \cite{Feng:2019tvb}. Instead, the authors gave
\bea
{\cal A}^{\epsilon,\W\epsilon}_{\rm{GR}}
(\{h_1,h_2\}\cup\pmb{H})=\sum_{f}\,\W\epsilon_{h_1}\cdot\W f_{h_f}\cdot {\cal B}_f+(\W\epsilon_{h_1}\cdot\W\epsilon_{h_2}){\cal A}^{\epsilon,\W\epsilon}_{\rm EYM}(h_1,h_2||\pmb{H})\,,
\eea
and
\bea
{\cal B}_f^\mu=-\sum_{\pmb{\alpha}\setminus h_f}\,(\W\epsilon_{h_2}\cdot\W f_{\alpha_m}\cdots\cdot\W f_{\alpha_2})^\mu{\cal A}^{\epsilon,\W\epsilon}_{\rm EYM}(h_1,h_f,\alpha_2,\cdots,\alpha_m,h_2||\pmb{H}\setminus\pmb{\alpha})\,.
\eea
To achieve the result \eref{multi-trace-ori1}, one need to be careful about the sign. Since tensors $\W f_i^{\mu\nu}$ are antisymmetric,
we have
\bea
\W\epsilon_{h_1}\cdot\W f_{h_f}\cdot {\cal B}_f=-\W\epsilon_{h_1}\cdot\W f_{h_f}\cdot(-\W f_{\alpha_2})\cdots(-\W f_{\alpha_m})\cdot\W\epsilon_{h_2}\,.
\eea
Tensors $(-\W f_i^{\mu\nu})$ cause the factor $(-)^{|\pmb{\alpha}|}$ in \eref{multi-trace-ori1}.
}
\bea
{\cal A}^{\epsilon,\W\epsilon}_{\rm{GR}}
(\{h_1,h_2\}\cup\pmb{H})&=&\sum_{\pmb{\bar{\alpha}}}(-)^{|\pmb{\alpha}|}\hat{C}(\pmb{\bar{\alpha}}){\cal A}^{\epsilon,\W\epsilon}_{\rm EYM}(h_1,\alpha_1,\cdots,\alpha_m,h_2||\pmb{H}
\setminus\pmb{\alpha})\,,~~~~\label{multi-trace-ori1}
\eea
where
\bea
\hat{C}(\pmb{\bar{\alpha}})=\W\epsilon_{h_1}\cdot \W f_{\alpha_1}\cdot \W f_{\alpha_2}\cdots \W f_{\alpha_{m}}\cdot \W\epsilon_{h_2}\,.~~~~\label{defin-D}
\eea
One can apply the expansion \eref{defin-D} at the first step then use \eref{defin-C} recursively to achieve the expansion of GR amplitudes into KK basis in the ordered splitting formula.

Before ending this subsection, we emphasize that the derivation of these two recursive expansions by solving differential equations serves as the first step in the construction of the unified web for expansions.

\section{Recursive expansions for multi-trace EYM amplitudes}
\label{multi-trace EYM}

To construct the unified web for expansions, the second step is deriving expansions of amplitudes of multi-trace EYM, EM and BI in the ordered splitting formula, by applying differential operators on expansions of GR and single-trace EYM amplitudes. This procedure have been done for EM and BI amplitudes in literatures \cite{Feng:2019tvb,Hu:2019qdq}. In this section, we derive two types of recursive expansions for multi-trace EYM amplitudes, by applying differential operators on the recursive expansion of single-trace EYM amplitudes. The similar recursive expansions are proposed in \cite{Du:2017gnh} from the angle of CHY integrands. Our purpose is to emphasize that they can be derived only through knowledge in Table \ref{tab:unifying}.
With two types of recursive expansions, the expansion of multi-trace EYM amplitudes in the ordered splitting formula can be obtained directly.

\subsection{Type-I recursive expansion}

The Type-I recursive expansion is applied to multi-trace EYM amplitudes ${\cal A}^{\epsilon,\W\epsilon}_{\rm{EYM}}
(\pmb{\bar{1}}|\pmb{\bar{2}}|\pmb{\bar{3}}|\cdots|\pmb{\bar{r}}||\{h_1,\pmb{H}\})$ which contain at least one graviton $h_1$.
Throughout this section, the ordered set $\pmb{\bar{1}}$ for the first trace is given explicitly as $\{1,2,\cdots,n\}$.
The Type-I recursive expansion expresses ${\cal A}^{\epsilon,\W\epsilon}_{\rm{EYM}}
(1,2,\cdots,n|\pmb{\bar{2}}|\pmb{\bar{3}}|\cdots|\pmb{\bar{r}}||\{h_1,\pmb{H}\})$ as the combination of multi-trace EYM amplitudes
with less gravitons, and can be derived through differential operators.
We begin with the special case that the EYM amplitudes contain two traces and at least one graviton $h_1$, ${\cal A}^{\epsilon,\W\epsilon}_{\rm{EYM}}
(1,2,\cdots,n|\pmb{\bar{2}}||\{h_1,\pmb{H}\})$, which can be generated by acting the the trace operator ${\cal T}^{\W\epsilon}[\pmb{\bar{2}}]$ on single-trace EYM amplitudes ${\cal A}^{\epsilon,\W\epsilon}_{\rm{EYM}}
(1,2,\cdots,n||\{h_1,\pmb{H},\pmb{2}\})$, i.e.,
\bea
{\cal A}^{\epsilon,\W\epsilon}_{\rm{EYM}}
(1,2,\cdots,n|\pmb{\bar{2}}||\{h_1,\pmb{H}\})={\cal T}^{\W\epsilon}[\pmb{\bar{2}}]{\cal A}^{\epsilon,\W\epsilon}_{\rm{EYM}}
(1,2,\cdots,n||\{h_1,\pmb{H},\pmb{2}\})\,.
\eea
The expansion
of ${\cal A}^{\epsilon,\W\epsilon}_{\rm{EYM}}
(1,2,\cdots,n|\pmb{\bar{2}}||\{h_1,\pmb{H}\})$ to pure YM amplitudes can be obtained by applying the operator ${\cal T}^{\W\epsilon}[\pmb{\bar{2}}]$ on two sides of the expansion of single-trace EYM amplitudes
\bea
{\cal A}^{\epsilon,\W\epsilon}_{\rm{EYM}}
(1,2,\cdots,n||\{h_1,\pmb{H},\pmb{2}\})&=&\sum_{\pmb{\bar{\alpha}}}\,C(\pmb{\bar{\alpha}}){\cal A}^{\epsilon,\W\epsilon}_{\rm EYM}(1,\{2,\cdots,n-1\}\shuffle\{\pmb{\bar{\alpha}},h_1\},n||\{\pmb{H},\pmb{2}\}
\setminus\pmb{\alpha})\,.~~~~\label{multi-trace-ori}
\eea

To analyse the effect of ${\cal T}^{\W\epsilon}[\pmb{\bar{2}}]$, one need to write down the explicit formula of this trace operator.
Suppose $\pmb{2}$ is a length-$r$ set
and $\pmb{\bar{2}}=\{\beta_1,\beta_2,\cdots,\beta_r\}$, the formula of ${\cal T}^{\W\epsilon}[\pmb{\bar{2}}]$ have several different equivalent choices,
for example both of
\bea
{\cal T}^{\W\epsilon}[\pmb{\bar{2}}]={\cal T}^{\W\epsilon}[\beta_1\beta_r]{\cal T}^{\W\epsilon}_{\beta_1\beta_2\beta_r}
{\cal T}^{\W\epsilon}_{\beta_2\beta_3\beta_r}\cdots{\cal T}^{\W\epsilon}_{\beta_{r-2}\beta_{r-1}\beta_r}
\eea
and
\bea
{\cal T}^{\W\epsilon}[\pmb{\bar{2}}]={\cal T}^{\W\epsilon}[\beta_1\beta_r]{\cal T}^{\W\epsilon}_{\beta_1\beta_{r-1}\beta_r}
{\cal T}^{\W\epsilon}_{\beta_1\beta_{r-2}\beta_{r-1}}\cdots{\cal T}^{\W\epsilon}_{\beta_{1}\beta_{2}\beta_3}
\eea
are correct. When applying ${\cal T}^{\W\epsilon}[\pmb{\bar{2}}]$ on the RHS of \eref{multi-trace-ori}, one need to ensure that the operators
act on different terms take the same formula. Thus, we fix the formula to be
\bea
{\cal T}^{\W\epsilon}[\pmb{\bar{2}}]={\cal T}^{\W\epsilon}[\beta_1\beta_r]{\cal T}^{\W\epsilon}_{\beta_1\beta_2\beta_r}
{\cal T}^{\W\epsilon}_{\beta_2\beta_3\beta_r}\cdots{\cal T}^{\W\epsilon}_{\beta_{r-2}\beta_{r-1}\beta_r}\,.~~~~\label{choice-insertion}
\eea
With such choice, one can discuss the expansion of double-trace amplitudes ${\cal A}^{\epsilon,\W\epsilon}_{\rm{EYM}}
(1,2,\cdots,n|\pmb{\bar{2}}||\{h_1,\pmb{H}\})$ by two cases.

\begin{itemize}
\item First case, $\pmb{2}\cap\pmb{\alpha}=\emptyset$.
\end{itemize}
For this case, the typical term in the summation on the RHS of \eref{multi-trace-ori} takes the formula
\bea
C(\pmb{\bar{\alpha}}){\cal A}^{\epsilon,\W\epsilon}_{\rm EYM}(1,\{2,\cdots,n-1\}\shuffle\{\pmb{\bar{\alpha}},h_1\},n||\{\pmb{H}\setminus\pmb{\alpha},\pmb{2}\}\,,
\eea
and will be turned into
\bea
C(\pmb{\bar{\alpha}}){\cal A}^{\epsilon,\W\epsilon}_{\rm EYM}(1,\{2,\cdots,n-1\}\shuffle\{\pmb{\bar{\alpha}},h_1\},n|\pmb{\bar{2}}||\pmb{H}\setminus\pmb{\alpha})
\eea
under the action of ${\cal T}^{\W\epsilon}[\pmb{\bar{2}}]$.

\begin{itemize}
\item Second case, $\pmb{2}\cap\pmb{\alpha}=\pmb{\phi}$.
\end{itemize}
Suppose $\pmb{2}=\pmb{\phi}\cup\pmb{\chi}$, where $\pmb{\phi}\subseteq\pmb{\alpha}$ and $\pmb{\chi}\cap\pmb{\alpha}=\emptyset$.
We first focus on the operator ${\cal T}^{\W\epsilon}[\beta_1\beta_r]$ in \eref{choice-insertion}. In general, one can find four configurations which are:
\bea
& &1:~\beta_1,\beta_r\in\pmb{\phi}\,,~~~~~~~~~~~~~~~~~~2:~ \beta_1,\beta_r\in\pmb{\chi}\,,\nn
& &3:~\beta_1\in\pmb{\phi}\,,~\beta_r\in\pmb{\chi}\,,~~~~~~~~~~~4:~\beta_1\in\pmb{\chi}\,,~\beta_r\in\pmb{\phi}\,.
\eea
However, for two cases in the second line, $\W\epsilon_{\beta_1}$ and $\W\epsilon_{\beta_r}$ do not contract with each other, therefore
the corresponding terms in \eref{multi-trace-ori} are annihilated by the operator ${\cal T}^{\W\epsilon}[\beta_1\beta_r]$. Thus, we only need to consider two cases, $\beta_1,\beta_r\in\pmb{\phi}$ and $\beta_1,\beta_r\in\pmb{\chi}$. Actually, the second one can also be excluded, as will be seen later.
For later convenience, we denote
\bea
\pmb{\alpha}\setminus\pmb{\phi}=\pmb{h},~~~~\pmb{2}\setminus\pmb{\phi}=\pmb{\chi}\,.~~~~\label{denote1}
\eea

For each term on the RHS in \eref{multi-trace-ori} with $\beta_1,\beta_r\in\pmb{\phi}$, the coefficient $C(\pmb{\bar{\alpha}})$ bears the action of ${\cal T}^{\W\epsilon}[\beta_1\beta_r]$.
If $C(\pmb{\bar{\alpha}})$ will not be annihilated by ${\cal T}^{\W\epsilon}[\beta_1\beta_r]$,
it must contain $(\W\epsilon_{\beta_1}\cdot\W\epsilon_{\beta_r})$. Due to the definition of $C(\pmb{\bar{\alpha}})$ in
\eref{defin-C}, this requirement indicates that $\beta_1$ and $\beta_r$ appear at nearby positions in $\pmb{\bar{\alpha}}$,
i.e., the ordered set $\pmb{\bar{\alpha}}$ takes the form $\{\cdots,\beta_r,\beta_1,\cdots\}$ or $\{\cdots,\beta_1,\beta_r,\cdots\}$.
Let us focus on the first case. The second case can be discussed similarly. For $\pmb{\bar{\alpha}}=\{\cdots,\beta_r,\beta_1,\cdots\}$,
the tensor $(\W f_{\beta_1}\cdot \W f_{\beta_r})^{\mu\nu}$ in $C(\pmb{\bar{\alpha}})$ is turned to $-k_{\beta_1}^\mu k_{\beta_r}^\nu$ by the operator ${\cal T}^{\W\epsilon}[\beta_1\beta_r]$, thus
\bea
{\cal T}^{\W\epsilon}[\beta_1\beta_r]C(\pmb{\bar{\alpha}})=-\big(\W\epsilon_{h_1}\cdot\W f_{\alpha_1}\cdots\W f_{\alpha_k}\cdot k_{\beta_1}\big)
\big(k_{\beta_r}\cdot\W f_{\alpha_{k+3}}\cdots\W f_{\alpha_m}\cdot Y_{\alpha_m}\big)\,.~~~~\label{C1}
\eea
Notice that the momenta $k_{\beta_1}^\mu$ and $k_{\beta_r}^\mu$ will not appear in any other place in
${\cal T}^{\W\epsilon}[\beta_1\beta_r]C(\pmb{\bar{\alpha}})$. In other words, $k_{\beta_1}^\mu$ appears once and only once, and so does $k_{\beta_r}^\nu$.

After applying ${\cal T}^{\W\epsilon}[\beta_1\beta_r]$, the remaining part of the trace operator ${\cal T}^{\W\epsilon}[\pmb{\bar{2}}]$
in \eref{choice-insertion} is consisted by insertion operators as
\bea
{\cal T}^{\W\epsilon}[\pmb{\bar{2}}]/{\cal T}^{\W\epsilon}[\beta_1\beta_r]={\cal T}^{\W\epsilon}_{\beta_1\beta_2\beta_r}
{\cal T}^{\W\epsilon}_{\beta_2\beta_3\beta_r}\cdots{\cal T}^{\W\epsilon}_{\beta_{r-2}\beta_{r-1}\beta_r}\,.
\eea
An insertion operator ${\cal T}^{\W\epsilon}_{\beta_{i-1}\beta_i\beta_r}$ acts on ${\cal T}^{\W\epsilon}[\beta_1\beta_r]C(\pmb{\bar{\alpha}})$
if $\beta_i\in\pmb{\phi}$, since $\W\epsilon_{\beta_i}$ is included in $\W f^{\mu\nu}_{\beta_i}$ in ${\cal T}^{\W\epsilon}[\beta_1\beta_r]C(\pmb{\bar{\alpha}})$. If $\beta_i\in\pmb{\chi}$, it acts on ${\cal A}^{\epsilon,\W\epsilon}_{\rm EYM}(1,\{2,\cdots,n-1\}\shuffle\{\pmb{\bar{\alpha}},h_1\},n||\{\pmb{H},\pmb{2}\}\setminus\pmb{\alpha})$.
To see the effect of these insertion operators clearly, one can expand them via the definition
\bea
{\cal T}^{\W\epsilon}_{\beta_{k-1}\beta_k\beta_r}={\partial\over\partial(\W\epsilon_{\beta_k}\cdot k_{\beta_{k-1}})}-{\partial\over\partial(\W\epsilon_{\beta_k}\cdot k_{\beta_r})}\,.
\eea
Then the operator ${\cal T}^{\W\epsilon}[\pmb{\bar{2}}]/{\cal T}^{\W\epsilon}[\beta_1\beta_r]$ is divided into several partial operators.
In each partial operator, an original operator ${\cal T}^{\W\epsilon}_{\beta_{k-1}\beta_k\beta_r}$ contributes one of to components ${\partial/\partial(\W\epsilon_{\beta_k}\cdot k_{\beta_{k-1}})}$ and $-{\partial/\partial(\W\epsilon_{\beta_k}\cdot k_{\beta_r})}$.
For insertion operators which act on ${\cal T}^{\W\epsilon}[\beta_1\beta_r]C(\pmb{\bar{\alpha}})$, the non-vanishing contributions correspond to partial operators that at most one original operator ${\cal T}^{\W\epsilon}_{\beta_{j-1}\beta_j\beta_r}$ contributes $-\partial/\partial(k_{\beta_r}\cdot\W\epsilon_{\beta_j})$,
while all other operators contribute $\partial/\partial(k_{\beta_{k-1}}\cdot\W\epsilon_{\beta_k})$,
due to the fact that $k_{\beta_r}$ appears only once in ${\cal T}^{\W\epsilon}[\beta_1\beta_r]C(\pmb{\bar{\alpha}})$. This observation leads to an important conclusion:
if one term on the RHS in \eref{multi-trace-ori} will not be annihilated by the trace operator ${\cal T}^{\W\epsilon}[\pmb{\bar{2}}]$, elements in the set $\pmb{\phi}$ should concentrate together in $\pmb{\bar{\alpha}}$, appear as
\bea
\pmb{\bar{\alpha}}=\{\alpha_m,\cdots,\alpha_{g+l+l'-j+2},\pmb{\bar{\phi}},\alpha_g,\cdots,\alpha_1\},~~~~{\rm with}~
\pmb{\bar{\phi}}=\{\beta_{l'},\beta_{l'-1},\cdots,\beta_j,\beta_r,\beta_1,\beta_2,\cdots,\beta_{l}\}\,,~~~~\label{phi1}
\eea
where $1\leq l<l'\leq r$,
and $l+1\leq j\leq l'$. More explicitly, elements in $\pmb{\bar{2}}$ are separated as
\bea
\pmb{\bar{2}}=\{\pmb{2}^1_l,\pmb{2}^{l+1}_{j-1},\pmb{2}^{j}_{l'},\pmb{2}^{l'+1}_{r-1},\beta_r\}\,,
\eea
and the ordered set $\pmb{\bar{\phi}}$ is given as
\bea
\pmb{\bar{\phi}}=\{\pmb{\W2}_{j}^{l'},\beta_r,\pmb{2}^1_l\}\,.
\eea
Here we have introduced
\bea
\pmb{2}^a_b=\{\beta_a,\beta_{a+1},\cdots,\beta_{b-1},\beta_b\}\,,~~~~~~\pmb{\W2}^a_b=\{\beta_a,\beta_{a-1},\cdots,\beta_{b+1},\beta_b\}\,.
\eea
If $a=b$, sets $\pmb{2}^a_b$ and $\pmb{\W2}^a_b$ are empty.

Now we explain the reason. We have demonstrated that the insertion operator
${\cal T}^{\W\epsilon}_{\beta_{i-1}\beta_i\beta_r}$ acts on
${\cal T}^{\W\epsilon}[\beta_1\beta_2]C(\pmb{\bar{\alpha}})$ if $\beta_i\in\pmb{\alpha}$. To continue, we use the observation that for insertion operators act on ${\cal T}^{\W\epsilon}[\beta_1\beta_2]C(\pmb{\bar{\alpha}})$, at most one of them contributes
$-\partial/\partial(k_{\beta_r}\cdot\W\epsilon_{\beta_j})$.
Suppose ${\cal T}^{\W\epsilon}_{\beta_{i-1}\beta_i\beta_r}$
contributes $\partial/\partial(k_{\beta_{i-1}}\cdot\W\epsilon_{\beta_i})$,
it requires the existence of $(k_{\beta_{i-1}}\cdot\W\epsilon_{\beta_i})$.
Thus if ${\cal T}^{\W\epsilon}[\beta_1\beta_2]C(\pmb{\bar{\alpha}})$ can survive under the action of $\partial/\partial(k_{\beta_{i-1}}\cdot\W\epsilon_{\beta_i})$, it should contain $(\W f_{\beta_{i-1}}\cdot \W f_{\beta_i})^{\mu\nu}$ or $(\W f_{\beta_{i}}\cdot \W f_{\beta_{i-1}})^{\mu\nu}$, since $k_{\beta_{i-1}}^\mu$ can not occur in $Y_{\alpha_m}^\mu$. Consequently, the operator $\partial/\partial(k_{\beta_{i-1}}\cdot\W\epsilon_{\beta_i})$ selects the ordering $\pmb{\bar{\alpha}}=\{\cdots,\beta_i,\beta_{i-1},\cdots\}$, or $\pmb{\bar{\alpha}}=\{\cdots,\beta_{i-1},\beta_{i},\cdots\}$, and turns $(\W f_{\beta_{i-1}}\cdot \W f_{\beta_i})^{\mu\nu}$ to $(-\W\epsilon_{\beta_{i-1}}^\mu)(-k_{\beta_i}^\nu)$, or $(\W f_{\beta_{i}}\cdot \W f_{\beta_{i-1}})^{\mu\nu}$ to $k_{\beta_{i}}^\mu\W\epsilon_{\beta_i-1}^\nu$. This conclusion shows $\beta_{i-1}\in\pmb{\alpha}$, thus the operator ${\cal T}^{\W\epsilon}_{\beta_{i-2}\beta_{i-1}\beta_r}$ also acts on ${\cal T}^{\W\epsilon}[\beta_1\beta_r]C(\pmb{\bar{\alpha}})$. Repeat the same argument,
one can conclude that if the operator ${\cal T}^{\W\epsilon}_{\beta_{i-2}\beta_{i-1}\beta_r}$
contributes $\partial/\partial(k_{\beta_{i-2}}\cdot\W\epsilon_{\beta_{i-1}})$, it would select $\pmb{\bar{\alpha}}=\{\cdots,\beta_i,\beta_{i-1},\beta_{i-2},\cdots\}$, or $\pmb{\bar{\alpha}}=\{\cdots,\beta_{i-2},\beta_{i-1},\beta_{i},\cdots\}$, and give
\bea
& &{\partial\over\partial(k_{\beta_{i-2}}\cdot\W\epsilon_{\beta_{i-1}})}{\partial\over\partial(k_{\beta_{i-1}}\cdot\W\epsilon_{\beta_{i}})}
(\W f_{\beta_{i-2}}\cdot \W f_{\beta_{i-1}}\cdot \W f_{\beta_i})^{\mu\nu}=(-)^3\W\epsilon_{\beta_{i-2}}^\mu k_{\beta_i}^\nu\,,\nn
& &{\partial\over\partial(k_{\beta_{i-2}}\cdot\W\epsilon_{\beta_{i-1}})}{\partial\over\partial(k_{\beta_{i-1}}\cdot\W\epsilon_{\beta_{i}})}
(\W f_{\beta_{i}}\cdot \W f_{\beta_{i-1}}\cdot \W f_{\beta_{i-2}})^{\mu\nu}=k_{\beta_i}^\mu\W\epsilon_{\beta_{i-2}}^\nu\,.
\eea
The conclusion $\beta_{i-2}\in\pmb{\alpha}$ allows us to repeat the same manipulation again.
Since the existence of $(\W f_{\alpha_l}\cdot k_{\beta_1})^\mu$ and $(k_{\beta_{r}}\cdot \W f_{\alpha_{l+3}})^\mu$ in \eref{C1}, this recursive pattern ends with two cases, one is
$\partial/\partial(k_{\beta_1}\cdot\W\epsilon_{\beta_2})$ in ${\cal T}^{\W\epsilon}_{\beta_1\beta_2\beta_r}$, another one is
$-\partial/\partial(k_{\beta_r}\cdot\W\epsilon_{\beta_j})$ in ${\cal T}^{\W\epsilon}_{\beta_{j-1}\beta_j\beta_r}$, with some $\beta_j\in\pmb{\alpha}$ and $j\neq 1,r$.
For the first case, we have $\pmb{\bar{\alpha}}=\{\cdots,\beta_r,\beta_1,\beta_2,\cdots,\beta_i,\cdots\}$,
and ${\cal T}^{\W\epsilon}[\beta_1\beta_r]C(\pmb{\bar{\alpha}})$ is turned to
\bea
-\big(\W\epsilon_{h_1}\cdot \W f_{\alpha_1}\cdots \W f_{\alpha_{k-i+1}}\cdot k_{\beta_i}\big)
\big(k_{\beta_r}\cdot \W f_{\alpha_{k+3}}\cdots \W f_{\alpha_m}\cdot Y_{\alpha_m}\big)\,.
\eea
For the second case we get $\pmb{\bar{\alpha}}=\{\cdots,\beta_{i},\beta_{i-1},\cdots,\beta_j,\beta_r,\beta_1,\cdots\}$,
and ${\cal T}^{\W\epsilon}[\beta_1\beta_r]C(\pmb{\bar{\alpha}})$ is turned to
\bea
(-)^{i-j}\big(\W\epsilon_{h_1}\cdot \W f_{\alpha_1}\cdots \W f_{\alpha_k}\cdot k_{\beta_1}\big)
\big(k_{\beta_i}\cdot \W f_{\alpha_{k+i-j+4}}\cdots \W f_{\alpha_m}\cdot Y_{\alpha_m}\big)\,.
\eea
The above discussion is correct for any $\beta_i\in\pmb{\alpha}$, thus we obtain the formulae \eref{phi1}.
It is straightforward to see that ${\cal T}^{\W\epsilon}[\beta_1\beta_r]C(\pmb{\bar{\alpha}})$ is transmuted to
\bea
(-)^{l'-j}(\W\epsilon_{h_1}\cdot \W f_{\alpha_1}\cdots \W f_{\alpha_g}\cdot k_{\beta_l})(k_{\beta_{l'}}\cdot \W f_{\alpha_{g+l+l'-j+3}}\cdots \W f_{\alpha_m}\cdot Y_{\alpha_m})\,.
\eea

To summarize the obtained color-orderings, we use notations in \eref{denote1} to arrive
\bea
& &{\cal T}^{\W\epsilon}[\pmb{\bar{2}}]\,\Big(\sum_{\substack{\pmb{\bar{\alpha}},\,\pmb{2}\cap\pmb{\alpha}\neq\emptyset,\\
\pmb{\bar{\alpha}}=\{\cdots,\beta_r,\beta_1,\cdots\}}}\
C(\pmb{\bar{\alpha}}){\cal A}^{\epsilon,\W\epsilon}_{\rm EYM}(1,\{2,\cdots,n-1\}\shuffle\{\pmb{\bar{\alpha}},h_1\},n||\{\pmb{H},\pmb{2}\}\setminus\pmb{\alpha})\Big)\nn
&=&{\cal T}^{\W\epsilon}[\pmb{\bar{2}}]\,\Big(\sum_{\substack{\pmb{\bar{\phi}},\,\pmb{\phi}\subseteq\pmb{2},\\
\pmb{\bar{\phi}}=\{\cdots,\beta_r,\beta_1,\cdots\}}}\sum_{\pmb{\bar{h}}}\,
C(\pmb{\bar{h}}{\shuffle}\underline{\pmb{\bar{\phi}}})
{\cal A}^{\epsilon,\W\epsilon}_{\rm EYM}(1,\{2,\cdots,n-1\}\shuffle\{\pmb{\bar{h}}{\shuffle}\underline{\pmb{\bar{\phi}}},h_1\},n||\{\pmb{H}\setminus\pmb{h},\pmb{\chi}\})\Big)\,.~~~~\label{sum-arrange}
\eea
Here we have introduced a new symbol $\underline{\pmb{\bar{A}}}$ to denote that the ordered set $\pmb{\bar{A}}$ appears as a single element
in the shuffle $\pmb{\bar{B}}{\shuffle}\underline{\pmb{\bar{A}}}$. For example
\bea
\{1,2\}{\shuffle}\underline{\pmb{\bar{A}}}=\{1,2,\pmb{\bar{A}}\}+\{1,\pmb{\bar{A}},2\}+\{\pmb{\bar{A}},1,2\}.
\eea

In the discussion mentioned above, the recursive pattern terminates at $(\W f_{\beta_2}\cdot k_{\beta_1})^\mu$,
or $(k_{\beta_r}\cdot \W f_{\beta_j})^\nu$. These two vectors in \eref{C1} arise from the assumption that
$\beta_1,\beta_r\in\pmb{\phi}$.
If $\beta_1,\beta_r\in\pmb{\chi}$, insertion operators ${\cal T}^{\W\epsilon}_{\beta_{i-1}\beta_i\beta_r}$ with $\beta_i\in\pmb{\phi}$ act on $C(\pmb{\bar{\alpha}})$ rather than ${\cal T}^{\W\epsilon}[\beta_1\beta_r]C(\pmb{\bar{\alpha}})$, and the recursive pattern can not be ended
before including all elements in $\pmb{2}$. This behavior violates the assumption $\beta_1,\beta_r\in\pmb{\chi}$. It implies that for the case
$\beta_1,\beta_r\in\pmb{\chi}$, the term on the RHS of \eref{multi-trace-ori} which will not vanish under the action of insertion operators does not
exist.

Insertion operators act on ${\cal T}^{\W\epsilon}[\beta_1\beta_2]C(\pmb{\bar{\alpha}})$ correspond to the subset $\pmb{2}^1_l\cup\pmb{2}^j_{l'}\cup\{\beta_r\}$,
thus the remaining insertion operators can be organized as
\bea
{\cal T}(\pmb{2}^{l+1}_{j-1})={\cal T}^{\W\epsilon}_{\beta_l\beta_{l+1}\beta_r}{\cal T}^{\W\epsilon}_{\beta_{l+1}\beta_{l+2}\beta_r}\cdots{\cal T}^{\W\epsilon}_{\beta_{j-2}\beta_{j-1}\beta_r}\,,
\eea
as well as
\bea
{\cal T}(\pmb{2}^{l'+1}_{r-1})={\cal T}^{\W\epsilon}_{\beta_{l'}\beta_{l'+1}\beta_r}{\cal T}^{\W\epsilon}_{\beta_{l'+1}\beta_{l'+2}\beta_r}\cdots{\cal T}^{\W\epsilon}_{\beta_{r-2}\beta_{r-1}\beta_r}\,,
\eea
which act on ${\cal A}^{\epsilon,\W\epsilon}_{\rm EYM}(1,\{2,\cdots,n-1\}\shuffle\{\pmb{\bar{h}}{\shuffle}\underline{\pmb{\bar{\phi}}},h_1\},n||\{\pmb{H}\setminus\pmb{h},\pmb{\chi}\})$ in \eref{sum-arrange}.
The operator ${\cal T}(\pmb{2}^{l'+1}_{r-1})$ can be interpreted as inserting the ordered set $\{\beta_{l'+1},\beta_{l'+2},\cdots,\beta_{r-1}\}$ between $\beta_{l'}$ and
$\beta_r$, while the operator ${\cal T}(\pmb{2}^{l+1}_{j-1})$, if be reorganized as
\bea
{\cal T}(\pmb{2}^{l+1}_{j-1})=(-)^{j-l-1}{\cal T}^{\W\epsilon}_{\beta_r\beta_{l+1}\beta_l}{\cal T}^{\W\epsilon}_{\beta_{r}\beta_{l+2}\beta_{l+1}}\cdots{\cal T}^{\W\epsilon}_{\beta_{r}\beta_{j-1}\beta_{j-2}}\,,
\eea
can be understood as inserting $\{\beta_{j-1},\beta_{j-2},\cdots,\beta_{l+1}\}$ between $\beta_r$ and $\beta_l$.
Thus we find
\bea
& &{\cal T}(\pmb{2}^{l+1}_{j-1}){\cal T}(\pmb{2}^{l'+1}_{r-1}){\cal A}^{\epsilon,\W\epsilon}_{\rm EYM}(1,\{2,\cdots,n-1\}\shuffle\{\pmb{\bar{h}}{\shuffle}\underline{\pmb{\bar{\phi}}},h_1\},n||\{\pmb{H}\setminus\pmb{h},\pmb{\chi}\})\nn
&=&(-)^{j-l-1}{\cal A}^{\epsilon,\W\epsilon}_{\rm EYM}(1,\{2,\cdots,n-1\}\shuffle\{\pmb{\bar{h}}{\shuffle}\underline{\pmb{\bar{\psi}}_1},h_1\},n||\pmb{H}\setminus\pmb{h})\,,
\eea
where
\bea
\pmb{\bar{\psi}}_1=\{\beta_{l'},\pmb{\W2}^{l'-1}_{j}\shuffle\pmb{2}^{l'+1}_{r-1},\beta_r,
\pmb{2}^1_{l-1}\shuffle\pmb{\W2}^{j-1}_{l+1},\beta_l\}\,.
\eea

Until now, we have studied the effect of insertion operators
for the ordering $\pmb{\bar{\alpha}}=\{\cdots,\beta_r,\beta_1,\cdots\}$.
Before simplify the result further, let us turn to the case $\pmb{\bar{\alpha}}=\{\cdots,\beta_1,\beta_{r},\cdots\}$.
The treatment is extremely similar.
One can find that elements in $\pmb{2}$ concentrate together as the ordered set
\bea
\pmb{\bar{\psi}}_2&=&\{\beta_{l},\pmb{\W2}^{l-1}_1\shuffle\pmb{2}^{l+1}_{j-1},\beta_r,
\pmb{2}^j_{l'-1}\shuffle\pmb{\W2}^{r-1}_{l'+1},\beta_{l'}\}\,,~~~~\label{psi2}
\eea
which is related to $\pmb{\bar{\psi}}_1$ by reversing the order.
For this case, we have
\bea
& &{\cal T}(\pmb{2}^{l+1}_{j-1}){\cal T}(\pmb{2}^{l'+1}_{r-1}){\cal A}^{\epsilon,\W\epsilon}_{\rm EYM}(1,\{2,\cdots,n-1\}\shuffle\{\pmb{\bar{h}}{\shuffle}\underline{\pmb{\bar{\phi}}},h_1\},n||\{\pmb{H}\setminus\pmb{h},\pmb{\chi}\})\nn
&=&(-)^{r-l'-1}{\cal A}^{\epsilon,\W\epsilon}_{\rm EYM}(1,\{2,\cdots,n-1\}\shuffle\{\pmb{\bar{h}}{\shuffle}\underline{\pmb{\bar{\psi}}_2},h_1\},n||\pmb{H}\setminus\pmb{h})\,,
\eea
and ${\cal T}^{\W\epsilon}[\beta_1\beta_r]C(\pmb{\alpha}^m)$ is turned to
\bea
(-)^{l}(\W\epsilon_{h_1}\cdot \W f_{\alpha_1}\cdots \W f_{\alpha_g}\cdot k_{\beta_{l'}})(k_{\beta_{l}}\cdot \W f_{\alpha_{g+l+l'-j+3}}\cdots \W f_{\alpha_m}\cdot Y_{\alpha_m})\,.
\eea

Above results for the current case $\pmb{2}\cap\pmb{\alpha}=\pmb{\phi}$ can be summarized as
\bea
& &{\cal T}^{\W\epsilon}[\pmb{\bar{2}}]\Big(\sum_{\substack{\pmb{\bar{\alpha}},\\\pmb{\alpha}=\pmb{\phi}\cup\pmb{h}}}\,C(\pmb{\bar{\alpha}}){\cal A}^{\epsilon,\W\epsilon}_{\rm EYM}
(1,\{2,\cdots,n-1\}\shuffle\{\pmb{\bar{\alpha}},h_1\},n||\{\pmb{H},\pmb{2}\}\setminus\pmb{\alpha})\Big)\nn
&=&\sum_{\pmb{\bar{h}}}\sum_{l}\sum_{l'}\sum_{j}\,(-)^{l'-l-1}C_{\beta_l,\beta_{l'}}(\pmb{\bar{h}}
{\shuffle}\underline{\pmb{\bar{\psi}}_1}){\cal A}^{\epsilon,\W\epsilon}_{\rm EYM}(1,\{2,\cdots,n-1\}\shuffle\{\pmb{\bar{h}}
{\shuffle}\underline{\pmb{\bar{\psi}}_1},h_1\},n||\pmb{H}\setminus\pmb{h})\nn
& &+\sum_{\pmb{\bar{h}}}\sum_{l}\sum_{l'}\sum_{j}\,(-)^{r-(l'-l)-1}C_{\beta_{l'},\beta_l}(\pmb{\bar{h}}
{\shuffle}\underline{\pmb{\bar{\psi}}_2}){\cal A}^{\epsilon,\W\epsilon}_{\rm EYM}(1,\{2,\cdots,n-1\}\shuffle\{\pmb{\bar{h}}
{\shuffle}\underline{\pmb{\bar{\psi}}_2},h_1\},n||\pmb{H}\setminus\pmb{h})\,.~~~~\label{4-terms}
\eea
The coefficient $C_{a,b}(\pmb{\bar{h}}
{\shuffle}\underline{\pmb{\bar{A}}})$ with $a,b\in\pmb{A}$ is defined as follows: suppose $\pmb{\bar{h}}
{\shuffle}\underline{\pmb{\bar{A}}}=\{\rho_{|\pmb{h}|+1},\rho_{|\pmb{h}|},\cdots,\rho_2,\rho_1\}$ (recall that $\pmb{\bar{A}}$ appears as a single element in $\pmb{\bar{h}}
{\shuffle}\underline{\pmb{\bar{A}}}$), $C_{a,b}(\pmb{\bar{h}}
{\shuffle}\underline{\pmb{\bar{A}}})$ is given as
\bea
C_{a,b}(\pmb{\bar{h}}
{\shuffle}\underline{\pmb{\bar{A}}})=\W\epsilon_{h_1}\cdot T_{\rho_1}\cdot T_{\rho_2}\cdots T_{\rho_{|\pmb{h}|+1}}\cdot Y_{\rho_{|\pmb{h}|+1}}\,,
~~~~\label{defin-C-2}
\eea
with
\bea
T_{\rho_i}^{\mu\nu}=\begin{cases}\displaystyle ~\W f_{h_k}^{\mu\nu}~~~~ & \rho_i=h_k\in\pmb{h}\,,\\
\displaystyle ~k_a^\mu k_b^\nu~~~~~ & \rho_i=\pmb{\bar{A}}\,.\end{cases}
\eea
If $\rho_{|\pmb{h}|+1}=\pmb{\bar{A}}$, $Y^\mu_{\rho_{|\pmb{h}|+1}}$ should be understood as $Y^\mu_b$.
Now we try to simplify \eref{4-terms}. We begin with the first term in \eref{4-terms}. By fixing $\pmb{\bar{h}}$, $l$ and $l'$, we get
\bea
& &(-)^{l'-l-1}C_{\beta_l,\beta_{l'}}(\pmb{\bar{h}}
{\shuffle}\underline{\pmb{\bar{\psi}}_1})\sum_{j=l+1}^{l'}\,{\cal A}^{\epsilon,\W\epsilon}_{\rm EYM}(1,\{2,\cdots,n-1\}\shuffle\{\pmb{\bar{h}}
{\shuffle}\underline{\pmb{\bar{\psi}}_1},h_1\},n||\pmb{H}\setminus\pmb{h})\nn
&=&(-)^{|\pmb{2}^{l}_{l'}|}C_{\beta_l,\beta_{l'}}(\pmb{\bar{h}}
{\shuffle}\underline{K^{\pmb{2}}_{\beta_{l'},\beta_l}}){\cal A}^{\epsilon,\W\epsilon}_{\rm EYM}(1,\{2,\cdots,n-1\}\shuffle\{\pmb{\bar{h}}
{\shuffle}\underline{K^{\pmb{2}}_{\beta_{l'},\beta_l}},h_1\},n||\pmb{H}\setminus\pmb{h})\,.~~~~\label{term-1}
\eea
Here $K^{\pmb{2}}_{\beta_{l'},\beta_l}$ stands for an ordered set which is given by reorganizing elements in $\pmb{\bar{2}}$. To define it, let us first use the cyclic symmetry to write $\pmb{\bar{2}}$ as
$\pmb{\bar{2}}=\{\beta_{l'},\pmb{2}^{l'+1}_{l-1},\beta_l,\pmb{2}^{l+1}_{l'-1}\}$,
then $K^{\pmb{2}}_{\beta_{l'},\beta_l}$ is given as
\bea
K^{\pmb{2}}_{\beta_{l'},\beta_l}=\{\beta_{l'},\pmb{2}^{l'+1}_{l-1}\shuffle\pmb{\W2}^{l'-1}_{l+1},\beta_l\}\,,~~~~\label{define-K}
\eea
where $\pmb{\W2}^{l'-1}_{l+1}$ is the reversing of $\pmb{2}^{l+1}_{l'-1}$.
When deriving \eref{term-1}, we have used the property of shuffle
\bea
\{\pmb{\bar{A}}_1,a,\pmb{\bar{A}}_2\}\shuffle\{b_1,\cdots,b_k\}\equiv\sum_{j=0}^k\,\{\pmb{\bar{A}}_1\shuffle\{b_1,\cdots,b_j\},a,
\pmb{\bar{A}}_2\shuffle\{b_{j+1},\cdots,b_k\}\}\,.
\eea
Consequently, the first term in \eref{4-terms} can be expressed as
\bea
\sum_{\pmb{\bar{h}}}\sum_l\sum_{l'}\,(-)^{|\pmb{2}^{l}_{l'}|}C_{\beta_l,\beta_{l'}}(\pmb{\bar{h}}
{\shuffle}\underline{K^{\pmb{2}}_{\beta_{l'},\beta_l}})\,{\cal A}^{\epsilon,\W\epsilon}_{\rm EYM}(1,\{2,\cdots,n-1\}\shuffle\{\pmb{\bar{h}}
{\shuffle}\underline{K^{\pmb{2}}_{\beta_{l'},\beta_l}},h_1\},n||\pmb{H}\setminus\pmb{h})\,.
\eea
Similar treatment holds for the second term in \eref{4-terms}, and provides
\bea
\sum_{\pmb{\bar{h}}}\sum_l\sum_{l'}\,(-)^{|\pmb{2}^{l'}_{l}|}C_{\beta_{l'},\beta_l}(\pmb{\bar{h}}
{\shuffle}\underline{K^{\pmb{2}}_{\beta_l,\beta_{l'}}})\,{\cal A}^{\epsilon,\W\epsilon}_{\rm EYM}(1,\{2,\cdots,n-1\}\shuffle\{\pmb{\bar{h}}
{\shuffle}\underline{K^{\pmb{2}}_{\beta_l,\beta_{l'}}},h_1\},n||\pmb{H}\setminus\pmb{h})\,.
\eea
Putting two terms together, we finally get
\bea
& &{\cal T}^{\W\epsilon}[\pmb{\bar{2}}]\Big(\sum_{\substack{\pmb{\bar{\alpha}},\\\pmb{\alpha}=\pmb{\phi}\cup\pmb{h}}}\,C(\pmb{\bar{\alpha}}){\cal A}^{\epsilon,\W\epsilon}_{\rm EYM}
(1,\{2,\cdots,n-1\}\shuffle\{\pmb{\bar{\alpha}},h_1\},n||\{\pmb{H},\pmb{2}\}\setminus\pmb{\alpha})\Big)\nn
&=&\sum_{\pmb{\bar{h}}}\sum_{a_2,b_2\in\pmb{2}}\,(-)^{|\pmb{2}^{a_2}_{b_2}|}C_{a_2,b_2}(\pmb{\bar{h}}
{\shuffle}\underline{K^{\pmb{2}}_{b_2,a_2}})\,{\cal A}^{\epsilon,\W\epsilon}_{\rm EYM}(1,\{2,\cdots,n-1\}\shuffle\{\pmb{\bar{h}}
{\shuffle}\underline{K^{\pmb{2}}_{b_2,a_2}},h_1\},n||\pmb{H}\setminus\pmb{h})\,.
\eea

Combining the results of two cases $\pmb{2}\cap\pmb{\alpha}=\emptyset$ and $\pmb{2}\cap\pmb{\alpha}\neq\emptyset$,
we find the full expansion of ${\cal A}^{\epsilon,\W\epsilon}_{EYM}(1,2,\cdots,n-1,n|\pmb{2}||\{h_1,\pmb{H}\})$,
which is given by
\bea
& &{\cal A}^{\epsilon,\W\epsilon}_{\rm EYM}(1,2,\cdots,n-1,n|\pmb{\bar{2}}||\{h_1,\pmb{H}\})\nn
&=&\sum_{\pmb{\bar{h}}}\,C(\pmb{\bar{h}})
{\cal A}^{\epsilon,\W\epsilon}_{\rm EYM}(1,\{2,\cdots,n-1\}\shuffle\{\pmb{\bar{h}},h_1\},n|\pmb{\bar{2}}||\pmb{H}\setminus\pmb{h})\nn
& &+\sum_{\pmb{\bar{h}}}\sum_{a_2,b_2\in\pmb{2}}\,(-)^{|\pmb{2}^{a_2}_{b_2}|}C_{a_2,b_2}(\pmb{\bar{h}}
{\shuffle}\underline{K^{\pmb{2}}_{b_2,a_2}})\,{\cal A}^{\epsilon,\W\epsilon}_{\rm EYM}(1,\{2,\cdots,n-1\}\shuffle\{\pmb{\bar{h}}
{\shuffle}\underline{K^{\pmb{2}}_{b_2,a_2}},h_1\},n||\pmb{H}\setminus\pmb{h})\,.
\eea

This expression can be generalized to multi-trace EYM amplitudes with arbitrary number of traces directly.
The general multi-trace EYM amplitudes ${\cal A}^{\epsilon,\W\epsilon}_{\rm{EYM}}
(1,2,\cdots,n|\pmb{\bar{2}}|\pmb{\bar{3}}|\cdots|\pmb{\bar{r}}||\{h_1,\pmb{H}\})$ can be created by applying trace operators
${\cal T}^{\W\epsilon}[\pmb{\bar{2}}],{\cal T}^{\W\epsilon}[\pmb{\bar{3}}],\cdots,{\cal T}^{\W\epsilon}[\pmb{\bar{r}}]$
on single-trace EYM amplitudes which can be expanded as
\bea
& &{\cal A}^{\epsilon,\W\epsilon}_{\rm{EYM}}
(1,2,\cdots,n||\{h_1,\pmb{H},\pmb{2},\pmb{3},\cdots,\pmb{r}\})\nn
&=&\sum_{\pmb{\bar{\alpha}}}\,C(\pmb{\bar{\alpha}}){\cal A}^{\epsilon,\W\epsilon}_{\rm EYM}(1,\{2,\cdots,n-1\}\shuffle\{\pmb{\bar{\alpha}},h_1\},n||\{\pmb{H},\pmb{2},\pmb{3},\cdots,\pmb{r}\}
\setminus\pmb{\alpha})\,.~~~~\label{multi-trace-EYM-gen}
\eea
Thus one can obtain the expansion of ${\cal A}^{\epsilon,\W\epsilon}_{\rm{EYM}}
(1,2,\cdots,n|\pmb{\bar{2}}|\pmb{\bar{3}}|\cdots|\pmb{\bar{r}}||\{h_1,\pmb{H}\})$
by performing trace operators on the RHS of \eref{multi-trace-EYM-gen}.
The calculation is exactly the same.
For each $\pmb{\phi}_i=\pmb{i}\cap\pmb{\alpha}$, one can find elements in $\pmb{\phi}_i$
appear in $\pmb{\bar{\alpha}}$ as $\pmb{\bar{\alpha}}=\{\cdots,\pmb{\bar{\phi}}_i,\cdots\}$,
and elements belong to $\pmb{i}/\pmb{\phi}_i$ are
inserted into $\pmb{\bar{\phi}}_i$.
Then, via the technique used for the double-trace case,
we obtain that
\bea
{\cal A}^{\epsilon,\W\epsilon}_{\rm EYM}(1,2,\cdots,n-1,n|\pmb{\bar{2}}|\pmb{\bar{3}}|\cdots|\pmb{\bar{r}}||\{h_1,\pmb{H}\})
&=&\sum_{\pmb{\bar{h}}}\sum_{\substack{\pmb{\bar{K}}(\pmb{\rm Tr}_s,a,b)\\\pmb{\rm Tr}_s\subset\pmb{\rm Tr}}}\Big[\,\underset{a_i,b_i\in\pmb{t}_i}{\widetilde{\sum}}\,\Big]^s_{i=1}\,A({\pmb{\bar{h}},\pmb{\bar{K}}(\pmb{\rm Tr}_s,a,b),h_1})\,,
\eea
where
\bea
A({\pmb{\bar{h}},\pmb{\bar{K}}(\pmb{\rm Tr}_s,a,b),h_1})=\W C_{a,b}(\pmb{\bar{h}}
{\shuffle}\pmb{\bar{K}})
{\cal A}^{\epsilon,\W\epsilon}_{\rm EYM}(1,\{2,\cdots,n-1\}\shuffle\{\pmb{\bar{h}}
{\shuffle}\pmb{\bar{K}}(\pmb{\rm Tr}_s,a,b),h_1\},n|\pmb{u}_1|\cdots|\pmb{u}_p||\pmb{H}\setminus\pmb{h})\,.
\eea

Explanations for notations are in order. The set ${\pmb{\rm Tr}}_s=\{\pmb{t_1},\pmb{t_2},\cdots,\pmb{t_s}\}$ is a subset of $\{\pmb{2},\pmb{3},\cdots,\pmb{r}\}$,
with ${\pmb{\rm Tr}}_s\cup\{\pmb{u}_1,\cdots,\pmb{u}_p\}=\{\pmb{2},\pmb{3},\cdots,\pmb{r}\}$.
Notice that $\pmb{\rm Tr}_s$ can be empty, which corresponds to the term
\bea
\sum_{\pmb{\bar{h}}}\,C(\pmb{\bar{h}})
{\cal A}^{\epsilon,\W\epsilon}_{\rm EYM}(1,\{2,\cdots,n-1\}\shuffle\{\pmb{\bar{h}},h_1\},n|\pmb{\bar{2}}|\pmb{\bar{3}}|\cdots|\pmb{\bar{r}}||\pmb{H}\setminus\pmb{h})\,.
\eea
The ordered set $\pmb{\bar{K}}(\pmb{\rm Tr}_s,a,b)$ is given by
\bea
\pmb{\bar{K}}(\pmb{\rm Tr}_s,a,b)=\{\underline{K^{\pmb{t_1}}_{b_1,a_1}},\underline{K^{\pmb{t_2}}_{b_2,a_2}},\cdots,\underline{K^{\pmb{t_s}}_{b_s,a_s}}\}\,,~~~~\label{defin-KK}
\eea
where $K^{\pmb{t_i}}_{b_i,a_i}$ is defined in \eref{define-K}. We want to emphasize two points. One is that
sets $K^{\pmb{t_i}}_{b_i,a_i}$ serve as individual elements in $\pmb{\bar{K}}(\pmb{\rm Tr}_s,a,b)$. Another one is,
$\pmb{\bar{K}}(\pmb{\rm Tr}_s,a,b)$ depends on the ordering of elements $K^{\pmb{t_i}}_{b_i,a_i}$, for example,
$\pmb{t_2}=\pmb{2},\pmb{t_3}=\pmb{3}$ and $\pmb{t_2}=\pmb{3},\pmb{t_3}=\pmb{2}$
do not provide the same $\pmb{\bar{K}}(\pmb{\rm Tr}_s,a,b)$. The definition of $\W C_{a,b}(\pmb{\bar{h}}
{\shuffle}\pmb{\bar{K}})$ is the generalization of \eref{defin-C-2}. Suppose
the length-$(|\pmb{h}|+s)$ set $\pmb{\bar{h}}
{\shuffle}\pmb{\bar{K}}(\pmb{\rm Tr}_s,a,b)$ is given as
\bea
\pmb{\bar{h}}
{\shuffle}\pmb{\bar{K}}(\pmb{\rm Tr}_s,a,b)=\{\rho_{|\pmb{h}|+s},\rho_{|\pmb{h}|+s-1},\cdots,\rho_2,\rho_1\}\,,
\eea
$\W C_{a,b}(\pmb{\bar{h}}
{\shuffle}\pmb{\bar{K}})$ corresponds to it is given as
\bea
\W C_{a,b}(\pmb{\bar{h}}
{\shuffle}\pmb{\bar{K}})=\W\epsilon_{h_1}\cdot T_{\rho_1}\cdot T_{\rho_2}\cdots T_{\rho_{|\pmb{h}|+s}}\cdot Y_{\rho_{|\pmb{h}|+s}}\,,
~~~~\label{defin-C-3}
\eea
with
\bea
T_{\rho_i}^{\mu\nu}=\begin{cases}\displaystyle ~\W f_{h_k}^{\mu\nu}~~~~ & \rho_i=h_k\in\pmb{h}\,,\\
\displaystyle ~k_{a_j}^\mu k_{b_j}^\nu~~~~~ & \rho_i=K^{\pmb{t_j}}_{b_j,a_j}\,.\end{cases}
\eea
If $\rho_{|\pmb{h}|+s}=K^{\pmb{t_j}}_{b_j,a_j}$, $Y^\mu_{\rho_{|\pmb{h}|+s}}$ is understood as $Y^\mu_{b_j}$.
The notation $\Big[\,\underset{a_i,b_i\in\pmb{t_i}}{\widetilde{\sum}}\,\Big]^s_{i=1}$ stands for
\bea
\Big[\,\underset{a_i,b_i\in\pmb{t}_i}{\widetilde{\sum}}\,\Big]^s_{i=1}\equiv
\underset{a_1,b_1\in\pmb{t}_1}{\widetilde{\sum}}\underset{a_2,b_2\in\pmb{t}_2}{\widetilde{\sum}}
\cdots\underset{a_s,b_s\in\pmb{t}_s}{\widetilde{\sum}}\,,~~~~\label{defin-sum}
\eea
where
\bea
\underset{a_i,b_i\in\pmb{t_i}}{\widetilde{\sum}}\equiv\sum_{{a_i,b_i\in\pmb{t_i}}}(-)^{|\pmb{t_i}^{a_i}_{b_i}|}\,.
\eea
%

\subsection{Type-II recursive expansion}

The Type-I recursive expansion applies to multi-trace EYM amplitudes whose external particles contain at least one graviton
so that one can always choose the fiducial graviton to be $h_1$. In this subsection, we turn to the case that all external particles are
gluons. For this case, we need to seek the new scheme of recursive expansion. The procedure of derivation bears strong similarity
with that in the previous subsection, thus we will omit many details.

We begin by considering the double-trace amplitude ${\cal A}^{\epsilon,\W\epsilon}_{\rm EYM}(1,2,\cdots,n-1,n|\pmb{\bar{2}}||\emptyset)$,
which can be created from the single-trace amplitudes ${\cal A}^{\epsilon,\W\epsilon}_{\rm EYM}(1,2,\cdots,n-1,n||\pmb{2})$
via the trace operator ${\cal T}^{\W\epsilon}[\pmb{\bar{2}}]$. To expand ${\cal A}^{\epsilon,\W\epsilon}_{\rm EYM}(1,2,\cdots,n-1,n||\pmb{2})$,
one need to choose a fiducial graviton in the set $\pmb{2}$. Let us denote this special graviton by $c_2$. Then we have
\bea
{\cal A}^{\epsilon,\W\epsilon}_{\rm EYM}(1,2,\cdots,n-1,n||\pmb{2})=\sum_{\pmb{\bar{\alpha}}}\,C(\pmb{\bar{\alpha}}){\cal A}^{\epsilon,\W\epsilon}_{\rm EYM}(1,\{2,\cdots,n-1\}\shuffle\{\pmb{\bar{\alpha}},c_2\},n||\pmb{2}
\setminus\{\pmb{\alpha},c_2\})\,,~~~~\label{multi-trace-ori2}
\eea
where $\pmb{\alpha}=\{\alpha_m,\cdots,\alpha_1\}$ with $m$ runs through $0$ to $|\pmb{2}|-1$, and
\bea
C(\pmb{\bar{\alpha}})=\W\epsilon_{c_2}\cdot \W f_{\alpha_1}\cdot \W f_{\alpha_2}\cdots \W f_{\alpha_m}\cdot Y_{\alpha_m}\,.
\eea
One can obtain the expansion of ${\cal A}^{\epsilon,\W\epsilon}_{\rm EYM}(1,2,\cdots,n-1,n|\pmb{\bar{2}}||\emptyset)$
by applying ${\cal T}^{\W\epsilon}[\pmb{2}]$ on two sides of \eref{multi-trace-ori2} simultaneously.

Similar as in the case $\pmb{2}\cap\pmb{\alpha}\neq\emptyset$ in the previous subsection, we first consider
$\beta_1,\beta_r\in\{\pmb{\alpha},c_2\}$, which yields $\{\pmb{\bar{\alpha}},c_2\}=\{\cdots,\beta_r,\beta_1,\cdots\}$, as well as
\bea
{\cal T}^{\W\epsilon}[\beta_1\beta_r]C(\pmb{\bar{\alpha}})=-(\W\epsilon_{c_2}\cdot \W f_{\alpha_1}\cdots \W f_{\alpha_l}\cdot k_{\beta_1})
(k_{\beta_r}\cdot \W f_{\alpha_{l+3}}\cdots \W f_{\alpha_m}\cdot Y_{\alpha_m})\,,
\eea
or $\{\pmb{\bar{\alpha}},c_2\}=\{\cdots,\beta_1,\beta_r,\cdots\}$, and correspondingly
\bea
{\cal T}^{\W\epsilon}[\beta_1\beta_r]C(\pmb{\bar{\alpha}})=-(\W\epsilon_{c_2}\cdot \W f_{\alpha_1}\cdots \W f_{\alpha_l}\cdot k_{\beta_r})
(k_{\beta_1}\cdot \W f_{\alpha_{l+3}}\cdots \W f_{\alpha_m}\cdot Y_{\alpha_m})\,.
\eea
Next, we repeat the manipulation in the previous subsection to get the sequences $\{\cdots,\beta_i,\beta_{i-1},\beta_{i-2},\cdots\}$ and $\{\cdots,\beta_{i-2},\beta_{i-1},\beta_i,\cdots\}$, which terminate when encountering $k_{\beta_1}^\mu$ or $k_{\beta_r}^\mu$.
The difference is, for the current case, all $\alpha_i$ satisfy $\alpha_i\in\pmb{2}$, thus one will find
$\{\pmb{\bar{\alpha}},c_2\}=\pmb{\W2}^{d_2}_{c_2}$ or $\{\pmb{\bar{\alpha}},c_2\}=\pmb{2}^{d_2}_{c_2}$, where $d_2=\alpha_m$, and $C(\pmb{\bar{\alpha}})$ is turned to $k_{d_2}\cdot Y_{d_2}$.

Cases $\beta_1,\beta_r\in\pmb{2}\setminus\{\pmb{\alpha},c_2\}$ or only one of them belongs to $\{\pmb{\alpha},c_2\}$ give no contribution, as can be
argued similar as in the previous subsection. Thus, the ordered set $\{\pmb{\alpha},c_2\}$ contains at least two elements $\beta_1$ and $\beta_r$.

Elements in the set $\pmb{2}\setminus\{\pmb{\alpha},c_2\}$ will be inserted into $\{\pmb{\bar{\alpha}},c_2\}$ by remaining insertion operators ${\cal T}^{\W\epsilon}_{\beta_{k-1}\beta_k\beta_r}$ with $\beta_k\in\pmb{2}\setminus\{\pmb{\alpha},c_2\}$. Again, the procedure is extremely similar
as that in the previous subsection. After these steps, we arrive the expansion
\bea
{\cal A}^{\epsilon,\W\epsilon}_{\rm EYM}(1,2,\cdots,n-1,n|\pmb{\bar{2}}||\emptyset)
=\underset{d_2\in\pmb{2}}{\widetilde{\sum}}\,(k_{d_2}\cdot Y_{d_2}){\cal A}^{\epsilon,\W\epsilon}_{\rm EYM}
(1,\{2,\cdots,n-1\}\shuffle \underline{K^{\pmb{2}}_{d_2,c_2}},n)\,.
\eea

Based on the above result, the expansion for the general multi-trace EYM amplitudes without external graviton can be obtained directly.
The treatment for other traces is the same as in the previous subsection since the fiducial graviton is already given. Thus we get
\bea
{\cal A}^{\epsilon,\W\epsilon}_{\rm EYM}(1,2,\cdots,n-1,n|\pmb{\bar{2}}|\pmb{\bar{3}}|\cdots|\pmb{\bar{r}}||\emptyset)
&=&\sum_{\substack{\pmb{\bar{K}}(\pmb{\rm Tr}_s,a,b)\\\pmb{\rm Tr}_s\subset\pmb{\rm Tr}/\pmb{2}}}\Big[\,\underset{a_i,b_i\in\pmb{t_i}}{\widetilde{\sum}}\,\Big]^s_{i=1}\underset{d_2\in\pmb{2}}{\widetilde{\sum}}\,
A({\pmb{\bar{K}}(\pmb{\rm Tr}_s,a,b),K^{\pmb{2}}_{d_2,c_2}})\,,
\eea
where
\bea
A({\pmb{\bar{K}}(\pmb{\rm Tr}_s,a,b),K^{\pmb{2}}_{d_2,c_2}})
&=&\hat{C}^{d_2}_{a,b}(\pmb{\bar{K}},K)
{\cal A}^{\epsilon,\W\epsilon}_{\rm EYM}(1,\{2,\cdots,n-1\}\shuffle\{\pmb{\bar{K}}(\pmb{\rm Tr}_s,a,b),\underline{K^{\pmb{2}}_{d_2,c_2}}\},n|\pmb{u}_1|\cdots|\pmb{u}_p||\emptyset)\,.~~~~\label{defin-A}
\eea
Here $\pmb{\rm Tr}_s=\{\pmb{t_1},\pmb{t_2},\cdots,\pmb{t_s}\}$ and $\pmb{\rm Tr}_s\cup\{\pmb{u}_1,\cdots,\pmb{u}_p\}=
\{\pmb{3},\cdots,\pmb{r}\}$. The ordered set $\pmb{\bar{K}}(\pmb{\rm Tr}_s,a,b)$ and the notation $\Big[\,\underset{a_i,b_i\in\pmb{t_i}}{\widetilde{\sum}}\,\Big]^s_{i=1}$ are defined in \eref{defin-KK} and \eref{defin-sum},
respectively. For $\pmb{\bar{K}}(\pmb{\rm Tr}_s,a,b)=\{\rho_s,\rho_{s-1},\cdots,\rho_1\}$, the coefficient $\hat{C}^{d_2}_{a,b}(\pmb{\bar{K}},K)$ is given as
\bea
\hat{C}^{d_2}_{a,b}(\pmb{\bar{K}},K)=k_{d_2}\cdot T_{\rho_1}\cdots T_{\rho_s}\cdot Y_{b_{\rho_s}}\,,
\eea
where $T^{\mu\nu}_{\rho_i}=k_{a_j}^\mu k_{b_j}^\nu$ if $\rho_i=K^{\pmb{t_j}}_{b_j,a_j}$.

Remarkably, $c_2$ is chosen as the fiducial graviton thus is fixed.
This is the reason we have the summation $\underset{d_2\in\pmb{2}}{\widetilde{\sum}}$ rather than $\underset{c_2,d_2\in\pmb{2}}{\widetilde{\sum}}$.
Also since $c_2$ is the fiducial graviton, we have the ordering $\{\pmb{\bar{K}}(\pmb{\rm Tr}_s,a,b),\underline{K^{\pmb{2}}_{d_2,c_2}}\}$ in \eref{defin-A}.

The type-II recursive expansion obtained in this paper is a little different from that in \cite{Du:2017gnh}, due to the different classification. In \cite{Du:2017gnh}, type-I
and type-II are classified by with or without the fiducial graviton. We classify them by with or without external graviton, which is the special case of the type-II expansion in \cite{Du:2017gnh}.

We have demonstrated that the type-I and type-II recursive expansions can be derived by using only differential operators.
Using two recursive expansions, one can expand all
multi-trace EYM amplitudes to KK basis of YM amplitudes in the ordered splitting formula, and extract coefficients of KK basis. The procedure is the same as that in \cite{Du:2017gnh},
thus we will not discuss this topic in the current paper. The algorithm of constructing coefficients of KK basis will be mentioned in the next section.

\section{Coefficients of KK basis for GR, EYM, EM and BI}
\label{coefficients}

After obtaining the expansions of GR, EYM, EM and BI in the ordered splitting formula, the third step is extracting coefficients of KK basis. For example, in the ordered splitting formula for the expansion of the single-trace EYM amplitude ${\cal A}^{\epsilon,\W\epsilon}_{\rm EYM}(1,2,3||\{4,5\})$, which is given as \footnote{The definition of the combinatory momentum $Z_i$ can be seen in the subsection \ref{algorithm}.}
\bea
{\cal A}^{\epsilon,\W\epsilon}_{\rm EYM}(1,2,3||\{4,5\})&=&\sum_{\shuffle}\,(\W\epsilon_4\cdot\W f_5\cdot Z_{5}){\cal A}^\epsilon_{\rm YM}(1,2\shuffle\{5,4\},3)\nn
& &+\sum_{\shuffle}\,(\W\epsilon_4\cdot Z_4)(\W\epsilon_5\cdot Z_5){\cal A}^\epsilon_{\rm YM}(1,2\shuffle\{4\}\shuffle\{5\},3)\,,
\eea
both terms ${\cal A}^\epsilon_{\rm YM}(1,2\shuffle\{5,4\},3)$ and ${\cal A}^\epsilon_{\rm YM}(1,2\shuffle\{4\}\shuffle\{5\},3)$ include the KK basis ${\cal A}^\epsilon_{\rm YM}(1,5,2,4,3)$. Thus, the coefficient of ${\cal A}^\epsilon_{\rm YM}(1,5,2,4,3)$ receives contributions from both of two terms.
In this section, we discuss how to get coefficients of KK basis. In the first subsection, we discuss the general relations among coefficients of KK basis for different theories, which are represented by differential operators.
In the second subsection, we introduce the systemic algorithms of constructing the coefficients for the KK basis with given color-ordering, which are already proposed in literatures \cite{Fu:2017uzt,Teng:2017tbo,Du:2017gnh,Feng:2019tvb,Hu:2019qdq}. We will explain that algorithms for other theories are related to the algorithm for GR.

\subsection{General relations among coefficients}
\label{rela-coe}

Before going to the evaluation of coefficients of KK basis, we first discuss the general relations among these coefficients. These coefficients can be extracted from the expansions in the ordered splitting formula. Since the expansions in the ordered splitting formula are derived through relations in Table \ref{tab:unifying}, one can claim that the same relations are satisfied by coefficients of KK basis. Let us demonstrate this consequence more explicitly.

Suppose we have obtained expansions of amplitudes of GR, EYM, EM and BI to KK basis, formally given as
\bea
& &{\cal A}^{\epsilon,\W\epsilon}_{\rm G}(\pmb{H}_n)=\sum_{\sigma}\,C^{\W\epsilon}_1(\sigma){\cal A}^{\epsilon}_{\rm YM}(1,\sigma_2,\sigma_3,\cdots,\sigma_{n-1},n)\,,\nn
& &{\cal A}^{\epsilon,\W\epsilon}_{\rm EYM}(\pmb{\bar{1}}|\pmb{\bar{2}}|\cdots|\pmb{\bar{r}}||\pmb{H}_{n-(|\pmb{1}|+|\pmb{2}|+\cdots+|\pmb{r}|)})=\sum_{\sigma}\,C^{\W\epsilon}_2(\sigma){\cal A}^{\epsilon}_{\rm YM}(1,\sigma_2,\sigma_3,\cdots,\sigma_{n-1},n)\,,\nn
& &{\cal A}^{\epsilon,\W\epsilon}_{\rm EMf}(\pmb{P}_{2m}||\pmb{H}_{n-2m})=\sum_{\sigma}\,C^{\W\epsilon}_{3a}(\sigma,\delta_{I_iI_j}){\cal A}^{\epsilon}_{\rm YM}(1,\sigma_2,\sigma_3,\cdots,\sigma_{n-1},n)\,,\nn
& &{\cal A}^{\epsilon,\W\epsilon}_{\rm EM}(\pmb{P}_{2m}||\pmb{H}_{n-2m})=\sum_{\sigma}\,C^{\W\epsilon}_{3b}(\sigma){\cal A}^{\epsilon}_{\rm YM}(1,\sigma_2,\sigma_3,\cdots,\sigma_{n-1},n)\,,\nn
& &{\cal A}^{\epsilon}_{\rm BI}(\pmb{P}_n)=\sum_{\sigma}\,C_4(\sigma){\cal A}^{\epsilon}_{\rm YM}(1,\sigma_2,\sigma_3,\cdots,\sigma_{n-1},n)\,,\nn
& &{\cal A}^{\epsilon}_{\rm YM}(i_1,i_2,\cdots,i_n)=\sum_{\sigma}\,C_5(\sigma){\cal A}^{\epsilon}_{\rm YM}(1,\sigma_2,\sigma_3,\cdots,\sigma_{n-1},n)\,,~~~~\label{expansion for 12345}
\eea
where the last line is nothing but the KK relation. Here we use $C^{\W\epsilon}_{3a}(\sigma,\delta_{I_iI_j})$ and $C^{\W\epsilon}_{3b}(\sigma)$ to distinguish coefficients for EMf and EM amplitudes. The summations of $\sigma$ are over all permutations of $(n-2)$ external legs in the group $S_{n-2}$. From Table \ref{tab:unifying}, we have following relations
\bea
& &{\cal A}^{\epsilon,\W\epsilon}_{\rm EYM}(\pmb{\bar{1}}|\pmb{\bar{2}}|\cdots|\pmb{\bar{r}}||\pmb{H}_{n-(|\pmb{1}|+|\pmb{2}|+\cdots+|\pmb{r}|)})={\cal T}^{\W\epsilon}[\pmb{\bar{1}}]\cdots{\cal T}^{\W\epsilon}[\pmb{\bar{r}}]{\cal A}^{\epsilon,\W\epsilon}_{\rm G}(\pmb{H}_n)\,,\nn
& &{\cal A}^{\epsilon,\W\epsilon}_{\rm EMf}(\pmb{P}_{2m}||\pmb{H}_{n-2m})={\cal T}^{\W\epsilon}_{{\cal X}_{2m}}{\cal A}^{\epsilon,\W\epsilon}_{\rm G}(\pmb{H}_n)\,,\nn
& &{\cal A}^{\epsilon,\W\epsilon}_{\rm EM}(\pmb{P}_{2m}||\pmb{H}_{n-2m})={\cal T}^{\W\epsilon}_{X_{2m}}{\cal A}^{\epsilon,\W\epsilon}_{\rm G}(\pmb{H}_n)\,,\nn
& &{\cal A}^{\epsilon}_{\rm BI}(\pmb{P}_n)={\cal L}^{\W\epsilon}{\cal T}^{\W\epsilon}[1n]{\cal A}^{\epsilon,\W\epsilon}_{\rm G}(\pmb{H}_n)\,,\nn
& &{\cal A}^{\epsilon}_{\rm YM}(i_1,i_2,\cdots,i_n)={\cal T}^{\W\epsilon}[i_1,i_2,\cdots,i_n]{\cal A}^{\epsilon,\W\epsilon}_{\rm G}(\pmb{H}_n)\,.
\eea
The key point is that the operators used above only modify coefficients in \eref{expansion for 12345}, since all of them are defined through polarization vectors $\W\epsilon_i$. Thus, one can conclude that
\bea
& &C^{\W\epsilon}_2(\sigma)={\cal T}^{\W\epsilon}[\pmb{\bar{1}}]\cdots{\cal T}^{\W\epsilon}[\pmb{\bar{r}}]C^{\W\epsilon}_1(\sigma)\,,\nn
& &C^{\W\epsilon}_{3a}(\sigma,\delta_{I_iI_j})={\cal T}^{\W\epsilon}_{{\cal X}_{2m}}C^{\W\epsilon}_1(\sigma)\,,\nn
& &C^{\W\epsilon}_{3b}(\sigma)={\cal T}^{\W\epsilon}_{X_{2m}}C^{\W\epsilon}_1(\sigma)\,,\nn
& &C_4(\sigma)={\cal L}^{\W\epsilon}{\cal T}^{\W\epsilon}[1n]C^{\W\epsilon}_1(\sigma)\,,\nn
& &C_5(\sigma)={\cal T}^{\W\epsilon}[i_1,i_2,\cdots,i_n]C^{\W\epsilon}_1(\sigma)\,.~~~~\label{action of operator}
\eea

There is a subtle point need to be clarified. Amplitudes of the KK basis are not independent to each other, due to the BCJ relations among them \cite{Bern:2008qj}. Thus coefficients in the formula \eref{expansion for 12345} are not unique. However, since all of these expansions are derived via relations provided by Table \ref{tab:unifying}, relations which are given in \eref{action of operator} still hold. In other words, one can get a variety of equivalent sets of coefficients related by the BCJ relations. Coefficients in each set satisfy relations summarized in \eref{action of operator}.

\subsection{Algorithms for evaluating coefficients}
\label{algorithm}

Coefficients of KK basis depend on external particles under consideration, as well as particular color-orderings of the basis. It is hard to find a general expression of coefficients which is correct for all cases. Instead, the systematic algorithms for evaluating them, which can be applied to any configuration of external particles and any color-ordering, can be provided. Algorithms for calculating coefficients which will be mentioned in this subsection have been proposed in literatures previously. We will introduce the algorithm for GR amplitudes first, then explain how algorithms for EYM, EM and BI amplitudes emergence from that for GR amplitudes.

Coefficients $C^{\W\epsilon}_1(\sigma)$ of KK basis for the expansion of GR amplitudes serve as the BCJ numerators for YM amplitudes. The corresponding algorithm is proposed in \cite{Fu:2017uzt}.
Assuming the set of gravitons is given as $\pmb{H}=\{1,2,\cdots,n\}$, and the KK basis is chosen to be ${\cal A}^{\epsilon}_{\rm YM}(1,\sigma_2,\cdots,\sigma_{n-1},n)$, with legs $1$ and $n$ are fixed at two ends. Our purpose is constructing the coefficient of the KK basis
with a given color-ordering. For later convenience, we denote the color-ordering
as $1\dot{<} \sigma_2\dot{<}\cdots\dot{<} \sigma_{n-1}\dot{<}n$. One also need to chose a reference ordering $n\prec j_2\prec\cdots\prec j_n$,
with $n$ is fixed at the lowest position. This reference ordering is denoted by $\pmb{\cal R}$.
The key step is constructing all correct ordered splittings consist with
the desired color-ordering, through the following procedure:
\begin{itemize}
\item At the first step, we construct all possible ordered subsets $\pmb{\bar{\alpha}}_0=\{1,\alpha^0_2,\cdots,\alpha^0_{|0|-1},n\}$, which satisfy two conditions, (1) $\pmb{{\alpha}}_0\subset\pmb{H}$, (2) $\alpha^0_2\dot{<}\alpha^0_3\dot{<}\cdots\dot{<}\alpha^0_{|0|-1}$, respecting to the color-ordering of the YM amplitude.
    Here $|i|$ stands for the length of the set $\pmb{\alpha}_i$.
    We call each ordered subset $\pmb{\bar{\alpha}}_0$ a root \footnote{Here we borrow the language from the framework of increasing spanning trees.}.
\item For each root $\pmb{\bar{\alpha}}_0$, we eliminate its elements in $\pmb{H}$ and $\pmb{\cal R}$, resulting in a reduced set $\pmb{H}\setminus\pmb{{\alpha}}_0$, and a reduce reference ordering $\pmb{\cal R}\setminus\pmb{{\alpha}}_0$. Suppose $R_1$ is the lowest element in the reduce reference ordering $\pmb{\cal R}\setminus\pmb{{\alpha}}_0$, we construct all possible ordered subsets $\pmb{\bar{\alpha}}_1$ as $\pmb{\bar{\alpha}}_1=\{\alpha_1^1,\alpha_2^1,\cdots,\alpha_{|1|-1}^1,R_1\}$, with $\alpha_1^1\dot{<}\alpha_2^1\dot{<}\cdots\dot{<}\alpha_{|1|-1}^1\dot{<}R_1$, regarding to the color-ordering.
\item By iterating the second step, one can construct $\pmb{\bar{\alpha}}_2,\pmb{\bar{\alpha}}_3,\cdots$, until $\pmb{\alpha}_0\cup\pmb{\alpha}_1\cup\cdots\cup\pmb{\alpha}_r=\pmb{H}$.
\end{itemize}
Each ordered splitting is given as an ordered set $\{\pmb{\bar{\alpha}}_0,\pmb{\bar{\alpha}}_1,\cdots,\pmb{\bar{\alpha}}_r\}$, where ordered sets
$\pmb{{\alpha}}_i$ serve as elements. For a given root $\pmb{\bar{\alpha}}_0$, an ordered set $\{\pmb{\bar{\alpha}}_1,\pmb{\bar{\alpha}}_2,\cdots,\pmb{\bar{\alpha}}_r\}$ is called a branch. Now we give the corresponding kinematic factors
for each ordered set $\pmb{\pmb{\alpha}}_i$. For a given ordered splitting, the root $\pmb{\bar{\alpha}}_0$ corresponds to the factor
\bea
(-)^{|\pmb{\alpha}_0|}(\W\epsilon_1\cdot \W f_{\alpha^0_2}\cdot \W f_{\alpha^0_3}\cdots \W f_{\alpha^0_{|0|-1}}\cdot\W\epsilon_n)\,.
\eea
Other ordered sets $\pmb{\pmb{\alpha}}_i$ with $i\neq 0$ correspond to
\bea
\W\epsilon_{R_i}\cdot \W f_{\alpha^i_{|i|-1}}\cdots \W f_{\alpha^i_2}\cdot \W f_{\alpha^i_1}\cdot Z_{\alpha^i_1}\,.
\eea
The combinatory momentum $Z_{\alpha^i_1}$ is the sum of momenta of external legs satisfying two conditions: (1) legs
at the LHS of $\alpha^i_1$ in the color-ordering, (2) legs belong to $\pmb{\bar{\alpha}}_j$ at the LHS of $\pmb{\bar{\alpha}}_i$
in the ordered splitting, i.e., $j<i$. The coefficient of the KK basis ${\cal A}^{\epsilon}_{\rm YM}(1,\sigma_2,\cdots,\sigma_{n-1},n)$
is the sum of contributions from all correct ordered splittings.

To illustrate this algorithm more clearly, let us consider an example, the coefficient of $4$-point YM amplitude ${\cal A}^{\epsilon}_{\rm YM}(1,3,2,4)$ with the color-ordering $1\dot{<}3\dot{<}2\dot{<}4$. The reference ordering is chosen to be $4\prec3\prec2\prec1$. The roots $\pmb{\bar{\alpha}}_0$ have following choices: $\{1,4\}$, $\{1,2,4\}$, $\{1,3,4\}$, $\{1,3,2,4\}$. The ordered set $\{1,2,3,4\}$ violates the color-ordering $3\dot{<}2$ therefore can not be a correct root. For the root $\pmb{\bar{\alpha}}_0=\{1,4\}$, the lowest element in the reduced reference ordering $3\prec2$ is $3$, then we can construct $\bar{\pmb{\alpha}}_1=\{3\}$ or $\pmb{\bar{\alpha}}_1=\{2,3\}$. However, the ordered set $\{2,3\}$ violates the color-ordering $3\dot{<}2$, therefore must be eliminated. Thus, we obtain the ordered splitting $\{\{1,4\},\{3\},\{2\}\}$ for the root $\{1,4\}$. Similarly, one can find $\{\{1,2,4\},\{3\}\}$, $\{\{1,3,4\},\{2\}\}$ and $\{\{1,3,2,4\}\}$ for other roots. After giving kinematic factors for these ordered splittings, the coefficient of ${\cal A}^{\epsilon}_{\rm YM}(1,3,2,4)$ is found to be
\bea
(\W\epsilon_1\cdot\W\epsilon_4)(\W\epsilon_3\cdot Z_3)(\W\epsilon_2\cdot Z_2)-(\W\epsilon_1\cdot\W f_2\cdot\W\epsilon_4)(\W\epsilon_3\cdot Z_3)-(\W\epsilon_1\cdot\W f_3\cdot\W\epsilon_4)(\W\epsilon_2\cdot Z_2)+(\W\epsilon_1\cdot\W f_3\cdot\W f_2\cdot\W\epsilon_4)\,.
\eea

Since the general relations among coefficients of KK basis in \eref{action of operator}, one can expect that algorithms for EYM, EM and BI are related to the algorithm for GR.
Now we explain these relations. We begin with EYM amplitudes. The corresponding algorithm for evaluating coefficients $C^{\W\epsilon}_2(\sigma)$ of KK basis was provided in \cite{Fu:2017uzt,Teng:2017tbo,Du:2017gnh}.
For the single-trace EYM amplitude ${\cal A}^{\epsilon,\W\epsilon}_{\rm EYM}(1,i_2,\cdots,i_{n-1},n||\pmb{H})$, the coefficient of the KK basis with the fixed color-ordering can be obtained as follows. Start with ordered splittings for the same color-ordering for the GR amplitude ${\cal A}^{\epsilon,\W\epsilon}_{\rm G}(\pmb{H}\cup\{1,i_2,\cdots,i_{n-1},n\})$, we collect all branches with the fixed root $\{1,i_2,\cdots,i_{n-1},n\}$.
The coefficient of the KK basis is the sum of contributions from these branches. For the multi-trace EYM amplitude ${\cal A}^{\epsilon,\W\epsilon}_{\rm EYM}(1,i_2,\cdots,i_{n-1},n|\pmb{\bar{2}}|\pmb{\bar{3}}|\cdots|\pmb{\bar{r}}||\pmb{H})$, one need to select correct
branches by two conditions: (1) Each ordered set $\pmb{\bar{j}}$ occurs in one of ordered subsets $\pmb{\bar{\alpha}}_i$ as a single element $K^{\pmb{j}}_{b_j,a_j}$ (or $K^{\pmb{2}}_{d_2,c_2}$), (2) the ordering of elements in $K^{\pmb{j}}_{b_j,a_j}$ consists with the color-ordering of the KK basis. The coefficient of KK basis is given by summing over contributions from selected branches.

Then we turn to EM amplitudes. The corresponding algorithm for calculating coefficients $C^{\W\epsilon}_{3a}(\sigma,\delta_{I_iI_j})$ and $C^{\W\epsilon}_{3b}(\sigma)$ was discussed in \cite{Hu:2019qdq}. For the EM amplitude ${\cal A}^{\epsilon,\W\epsilon}_{\rm EM}(\pmb{P}_{2m}||\pmb{H})$ which contains $2m$ external photons, photons are grouped into $m$ pairs as $\{(i_1,j_1),(i_2,j_2),\cdots(i_m,j_m)\}$, with $i_1<i_2<\cdots<i_m$ and $i_k<j_k$ for $\forall k$. Each proper set of pairs is called a partition. For each partition, one can start with ordered splittings for the desired color-ordering for the GR amplitude ${\cal A}^{\epsilon,\W\epsilon}_{\rm G}(\pmb{H}\cup\pmb{P}_{2m})$, and select ordered splittings by the condition each pair in the partition appears in one subset as a single element, i.e., two photons are at nearby positions. Then, for a selected ordered splitting, a subset $\pmb{\alpha}_0$ carries the original factor $(-)^{|0|}(\W\epsilon_{1}\cdot\W f_{\alpha^0_2}\cdots\W f_{\alpha^0_{|0|-1}}\cdot\W \epsilon_n)$, and subsets $\pmb{\alpha}_{i\neq0}$ carry $\W\epsilon_{\alpha_{R_i}}\cdot\W f_{\alpha^i_{|i|-1}}\cdots\W f_{\alpha^i_1}\cdot Z_{\alpha^i_1}$. The contribution from the subset $\pmb{\alpha}_i$ is obtained by selecting terms which contain all contractions $(\W\epsilon_{i_k}\cdot\W\epsilon_{j_k})$ with $i_k,j_k\in\pmb{\alpha}$ in the original factor, and turning each $(\W\epsilon_{i_k}\cdot\W\epsilon_{j_k})$
to $1$. The coefficient of KK basis with the given color-ordering is obtained by summing over contributions from selected ordered splittings, and summing over all allowed partitions.

For the BI amplitude ${\cal A}^{\epsilon}_{\rm BI}(\pmb{P}_{2m})$, there are two equivalent expressions of the coefficient $C_4(\sigma)$. The first one is provided in \cite{Feng:2019tvb}, which is given as $\prod_{l\in\pmb{P}_{2m}}\,(k_l\cdot X_l)$, where $X_l$ is defined as the sum of momenta for legs on the LHS of the leg $l$ in the color-ordering. To obtain another expression, one can start with ordered splittings for the given color-ordering for the GR amplitude ${\cal A}^{\epsilon,\W\epsilon}_{\rm G}(\pmb{P}_{2m})$, collect all branches having the root $\{1,n\}$ and satisfying the condition each ordered subset in branches has the even length. Then, for each subset $\pmb{\alpha}_i$ in a selected branch, we pick out the term $(-)^{|\pmb{\alpha}_i|\over 2}(\W\epsilon_{{R_i}}\cdot \W\epsilon_{\alpha^i_{|i|-1}})(k_{\alpha^i_{|i|-1}}\cdot k_{\alpha^i_{|i|-2}})\cdots(\W\epsilon_{\alpha^i_3}\cdot \W\epsilon_{\alpha^i_2})(k_{\alpha^i_2}\cdot k_{\alpha^i_1})$ in the original factor $\W\epsilon_{\alpha_{R_i}}\cdot\W f_{\alpha^i_{|i|-1}}\cdots\W f_{\alpha^i_1}\cdot Z_{\alpha^i_1}$,
and turn each $(\W\epsilon^i_a\cdot \W\epsilon^i_b)$ to $(k^i_a\cdot k^i_b)$, to obtain the contribution $(-)^{|\pmb{\alpha}_i|\over 2}(k_{{R_i}}\cdot k_{\alpha^i_{|i|-1}})(k_{\alpha^i_{|i|-1}}\cdot k_{\alpha^i_{|i|-2}})\cdots(k_{\alpha^i_3}\cdot k_{\alpha^i_2})(k_{\alpha^i_2}\cdot k_{\alpha^i_1})$. The coefficient of KK basis with the desired color-ordering is provided by summing over contributions from selected branches.

In this paper, the procedure leads to algorithms for evaluating coefficients is: (1) derive the expansion in the ordered splitting formula for the theory under consideration, (2) extract the coefficients from the expansion in the ordered splitting formula. This is the method used in literatures.
However, as pointed out above, algorithms for other theories can be generated from the algorithm for GR. Thus, an alternative procedure can be proposed naturally. One can derive the algorithm for GR at first, then obtain algorithms for other theories by applying relations in \eref{action of operator}.

As a simple example, let us derive the algorithm for evaluating coefficients $C^{\W\epsilon}_2(\sigma)$ for single trace EYM amplitudes, by using the relation
\bea
C^{\W\epsilon}_2(\sigma)={\cal T}^{\W\epsilon}[\pmb{\bar{1}}]C^{\W\epsilon}_1(\sigma)\,.
\eea
Without lose of generality, we assume the length-$m$ ordered set $\pmb{\bar{1}}$ is $\pmb{\bar{1}}=\{1,i_2,i_3,\cdots,i_{m-1},n\}$, and the formula of the trace operator is chosen to be
\bea
{\cal T}^{\W\epsilon}[\pmb{\bar{1}}]={\cal T}^{\W\epsilon}[1n]{\cal T}^{\W\epsilon}_{1i_2n}{\cal T}^{\W\epsilon}_{i_2i_3n}\cdots{\cal T}^{\W\epsilon}_{i_{m-2}i_{m-1}n}\,.
\eea
To simplify the calculation, we choose the reference ordering as $n\prec i_2\prec i_3\prec\cdots i_{m-1}\prec1$. Under such choices,
correct roots take the form $\pmb{\bar{\alpha}}_0=\{1,\alpha^0_2,\alpha^0_3,\cdots,\alpha^0_{|0|-1},n\}$ when expanding GR amplitudes.
Among these roots, only $\{1,n\}$ provides the contraction $(\W\epsilon_1\cdot\W\epsilon_n)$ which is required by the operator ${\cal T}^{\W\epsilon}[1n]$ in ${\cal T}^{\W\epsilon}[\pmb{\bar{1}}]$. Thus, under the action of ${\cal T}^{\W\epsilon}[\pmb{\bar{1}}]$, non-vanishing contributions arise from
ordered splittings with the root $\pmb{\bar{\alpha}}_0=\{1,n\}$. At this step, the factor $(\W\epsilon_1\cdot\W\epsilon_n)$ provided by the root is turned to $1$ by the operator ${\cal T}^{\W\epsilon}[1n]$. Then, in the reduced reference ordering $\pmb{R}\setminus\pmb{\alpha}_0$, the lowest element is $i_2$. The corresponding ordered subset $\pmb{\bar{\alpha}}_1$ has the form $\pmb{\bar{\alpha}}_1=\{\alpha^1_1,\alpha^1_2,\cdots,\alpha^1_{|1|-1},i_2\}$, which corresponds to the factor $\W\epsilon_{i_2}\cdot\W f_{\alpha^1_{|1|-1}}\cdots\W f_{\alpha^1_2}\cdot\W f_{\alpha^1_1}\cdot Z_{\alpha^1_1}$. Under the action of the operator ${\cal T}^{\W\epsilon}_{1i_2n}$, the survived term should include $(\W\epsilon_{i_2}\cdot k_1)$ or $(\W\epsilon_{i_2}\cdot k_n)$. In the factor $\W\epsilon_{i_2}\cdot\W f_{\alpha^1_{|1|-1}}\cdots\W f_{\alpha^1_2}\cdot\W f_{\alpha^1_1}\cdot Z_{\alpha^1_1}$, $k_1$ only appears in $Z_{\alpha^1_1}$, while $k_n$ does not appear. Thus, the non-zero contribution comes from the ordered subset $\pmb{\bar{\alpha}}_1=\{i_2\}$. At this step, the factor $\W\epsilon_{i_2}\cdot Z_{i_2}$ provided by the subset $\pmb{\bar{\alpha}}_1$ is turned to $1$. To continue, we construct the ordered subset $\pmb{\bar{\alpha}}_2=\{\alpha^2_1,\alpha^2_2,\cdots,\alpha^2_{|2|-1},i_3\}$, where $i_3$ is the lowest element in the reduced reference ordering $\pmb{R}\setminus\{\pmb{\alpha}_0\cup\pmb{\alpha}_1\}$. The operator ${\cal T}^{\W\epsilon}_{i_2i_3n}$ requires the existence of the contraction $(\W\epsilon_{i_3}\cdot k_{i_2})$ or $(\W\epsilon_{i_3}\cdot k_n)$. In the factor $\W\epsilon_{i_3}\cdot\W f_{\alpha^2_{|2|-1}}\cdots\W f_{\alpha^2_2}\cdot\W f_{\alpha^2_1}\cdot Z_{\alpha^2_1}$, $k_{i_2}$ appears in $Z_{\alpha^2_1}$ when $i_2\dot{<} i_3$ respecting to the given color-ordering, while $k_n$ doe not appear. Thus, if $i_2\dot{<} i_3$, the non-vanishing contribution corresponds to the ordered subset $\pmb{\bar{\alpha}}_2=\{i_3\}$. Otherwise, the coefficient of the KK basis with desired color-ordering is zero. At this step, the factor $\W\epsilon_{i_3}\cdot Z_{i_3}$ provided by the subset $\pmb{\bar{\alpha}}_2$ is turned to $1$. Repeat the manipulation of constructing $\pmb{\bar{\alpha}}_2$, one can find that for non-zero contributions to the coefficient, the correct ordered splittings must have the structure $\pmb{\bar{\alpha}}=\{\pmb{\bar{\alpha}}_0,\pmb{\bar{\alpha}}_1,\cdots,\pmb{\bar{\alpha}}_r\}$ with $\pmb{\bar{\alpha}}_0=\{1,n\}$ and $\pmb{\bar{\alpha}}_j=\{i_{j+1}\}$ for $\forall 1\leq j\leq m-2$, satisfying the condition $i_2\dot{<} i_3\dot{<}\cdots\dot{<}i_{m-1}$.
After applying the operator ${\cal T}^{\W\epsilon}[\pmb{\bar{1}}]$, ordered subsets $\pmb{\bar{\alpha}}_0$ and all of $\pmb{\bar{\alpha}}_j$ for $\forall 1\leq j\leq m-2$ contribute $1$. Thus the non-trivial contributions are provided by ordered subsets $\pmb{\bar{\alpha}}_{m-1},\pmb{\bar{\alpha}}_m,\cdots,\pmb{\bar{\alpha}}_r$. The ordered sets $\pmb{\bar{\alpha}}\setminus\{\pmb{\bar{\alpha}}_0,\cdots,\pmb{\bar{\alpha}}_{m-2}\}=\{\pmb{\bar{\alpha}}_{m-1},\pmb{\bar{\alpha}}_m,
\cdots,\pmb{\bar{\alpha}}_r\}$ are branches for the root $\{1,i_2,i_3,\cdots,i_{m-1},n\}$, and can be identified as building blocks for $C^{\W\epsilon}_2(\sigma)$. Notice that the root for branches $\{\pmb{\bar{\alpha}}_{m-1},\pmb{\bar{\alpha}}_m,\cdots,\pmb{\bar{\alpha}}_r\}$ which consists with the color-ordering has only one candidate, due to the condition $i_2\dot{<} i_3\dot{<}\cdots\dot{<}i_{m-1}$. Picking up factors $\W\epsilon_{R_i}\cdot \W f_{\alpha^i_{|i|-1}}\cdots \W f_{\alpha^i_2}\cdot \W f_{\alpha^i_1}\cdot Z_{\alpha^i_1}$ for subsets $\bar{\pmb{\alpha}}_i$ in these branches, the algorithm for evaluating $C^{\W\epsilon}_2(\sigma)$ is achieved. Consequently, the algorithm for single-trace EYM can be derived by applying general relations in \eref{action of operator} on the algorithm for GR. Algorithms for other theories can be discussed similarly.

\section{Unified web for expansions}
\label{unified web}

In this section, we construct the complete unified web for expansions which contains all theories in Table \ref{tab:unifying}, by applying differential operators. In the first subsection, we derive expansions of amplitudes of YM, YMS, sYMS, $\phi^4$, NLSM, BS, BI, exDBI, DBI, as well as SG, and give the complete unified web. This is the final step of constructing the unified web of expansions. In the second section, we unify all expansions in the web into the compact double copy formula, and discuss the duality between the web for expansions and the web for differential operators.

\subsection{Complete unified web}

In previous two sections, differential operators under consideration are defined through polarization vectors $\W\epsilon_i$. In this section, we consider operators defined via $\epsilon_i$. When acting on expansions in \eref{expansion for 12345}, these operators transmute amplitudes on both sides to amplitudes of other theories, without modifying coefficients. This manipulation gives expansions of amplitudes of other theories beyond those discussed previously.

First, we apply the operator ${\cal T}^\epsilon[i'_1,\cdots,i'_n]$ on two sides of each expansion in \eref{expansion for 12345}. Using relations
in Table \ref{tab:unifying}, we obtain expansions of amplitudes of YM, YMS, sYMS, $\phi^4$, NLSM and BAS to BAS amplitudes, which are given by
\bea
& &{\cal A}^{\W\epsilon}_{\rm YM}(i'_1,\cdots,i'_n)=\sum_{\sigma}\,C^{\W\epsilon}_1(\sigma){\cal A}_{\rm BAS}(1,\sigma_2,\cdots,\sigma_{n-1},n;i'_1,\cdots,i'_n)\,,\nn
& &{\cal A}^{\W\epsilon}_{\rm YMS}(\pmb{\bar{1}}|\cdots|\pmb{\bar{r}}||\pmb{G}_{n-(|\pmb{1}|+\cdots+|\pmb{r}|)};i'_1,\cdots,i'_n)
=\sum_{\sigma}\,C^{\W\epsilon}_2(\sigma){\cal A}_{\rm BAS}(1,\sigma_2,\cdots,\sigma_{n-1},n;i'_1,\cdots,i'_n)\,,\nn
& &{\cal A}^{\W\epsilon}_{\rm sYMS}(\pmb{S}_{2m}||\pmb{G}_{n-2m};i'_1,\cdots,i'_n)=\sum_{\sigma}\,C^{\W\epsilon}_{3a}(\sigma,\delta_{I_iI_j}){\cal A}_{\rm BAS}(1,\sigma_2,\cdots,\sigma_{n-1},n;i'_1,\cdots,i'_n)\,,\nn
& &{\cal A}_{\phi^4}(i'_1,\cdots,i'_n)=\sum_{\sigma}\,C_{3b}(\sigma){\cal A}_{\rm BAS}(1,\sigma_2,\cdots,\sigma_{n-1},n;i'_1,\cdots,i'_n)\,,\nn
& &{\cal A}_{\rm NLSM}(i'_1,\cdots,i'_n)=\sum_{\sigma}\,C_4(\sigma){\cal A}_{\rm BAS}(1,\sigma_2,\cdots,\sigma_{n-1},n;i'_1,\cdots,i'_n)\,,\nn
& &{\cal A}_{\rm BAS}(l_1,\cdots,l_n;i'_1,\cdots,i'_n)=\sum_{\sigma}\,C_5(\sigma){\cal A}_{\rm BAS}(1,\sigma_2,\cdots,\sigma_{n-1},n;i'_1,\cdots,i'_n)\,.~~~~\label{ex-12345-KK2}
\eea
In the above expansions, the basis are BAS amplitudes with $1$ and $n$ are fixed at two ends in one of two color-orderings. One can use
\bea
{\cal A}_{\rm BAS}(1,\sigma_2,\cdots,\sigma_{n-1},n;i'_1,i'_2,\cdots,i'_n)=\sum_{\sigma'}\,C_5(\sigma'){\cal A}_{\rm BAS}(1,\sigma_2,\cdots,\sigma_{n-1},n;1,\sigma'_2,\cdots,\sigma'_{n-1},n)
\eea
to expand amplitudes in \eref{ex-12345-KK2} to BAS amplitudes ${\cal A}_{\rm BAS}(1,\sigma_2,\cdots,\sigma_{n-1},n;1,\sigma'_2,\cdots,\sigma'_{n-1},n)$.
The obtained expansions are examples of the double copy formula which will be discussed in the next subsection.

Secondly, we apply the operator ${\cal L}^\epsilon\cdot{\cal T}^\epsilon[ab]$ on two sides of each expansion in \eref{expansion for 12345}. Using relations in Table \ref{tab:unifying}, we get expansions of amplitudes of BI, exDBI, DBI, SG and NLSM to NLSM amplitudes, which are given as
\bea
& &{\cal A}^{\W\epsilon}_{\rm BI}(\pmb{P}_n)=\sum_{\sigma}\,C^{\W\epsilon}_1(\sigma){\cal A}_{\rm NLSM}(1,\sigma_2,\cdots,\sigma_{n-1},n)\,,\nn
& &{\cal A}^{\W\epsilon}_{\rm exDBI}(\pmb{\bar{1}}|\cdots|\pmb{\bar{r}}||\pmb{P}_{n-(|\pmb{1}|+\cdots+|\pmb{r}|)})
=\sum_{\sigma}\,C^{\W\epsilon}_2(\sigma){\cal A}_{\rm NLSM}(1,\sigma_2,\cdots,\sigma_{n-1},n)\,,\nn
& &{\cal A}^{\W\epsilon}_{\rm DBI}(\pmb{S}_{2m}||\pmb{P}_{n-2m})=\sum_{\sigma}\,C^{\W\epsilon}_{3a}(\sigma,\delta_{I_iI_j}){\cal A}_{\rm NLSM}(1,\sigma_2,\cdots,\sigma_{n-1},n)\,,\nn
& &{\cal A}_{\rm SG}(\pmb{S}_n)=\sum_{\sigma}\,C_4(\sigma){\cal A}_{\rm NLSM}(1,\sigma_2,\cdots,\sigma_{n-1},n)\,,\nn
& &{\cal A}_{\rm NLSM}(l_1,\cdots,l_n)=\sum_{\sigma}\,C_5(\sigma){\cal A}_{\rm NLSM}(1,\sigma_2,\cdots,\sigma_{n-1},n)\,.~~~~\label{ex-12345-KK3}
\eea

Thirdly, we apply operators ${\cal T}^\epsilon[\pmb{\bar{1'}}]\cdots{\cal T}^\epsilon[\pmb{\bar{r'}}]$, ${\cal T}^\epsilon_{{\cal X}_{2m'}}$ and ${\cal T}^\epsilon_{X_n}$ on two sides of the expansion of GR amplitudes in \eref{expansion for 12345} to get
\bea
& &{\cal A}^{\W\epsilon,\epsilon}_{\rm EYM}(\pmb{\bar{1'}}|\cdots|\pmb{\bar{r'}}||\pmb{G}_{n-(|\pmb{1'}|+\cdots+|\pmb{r'}|)})
=\sum_{\sigma}\,C^{\W\epsilon}_1(\sigma){\cal A}^\epsilon_{\rm YMS}(\pmb{\bar{1'}}|\cdots|\pmb{\bar{r'}}||\pmb{G}_{n-(|\pmb{1'}|+\cdots+|\pmb{r'}|)};1,\sigma_2,\cdots,\sigma_{n-1},n)\nn
& &{\cal A}^{\W\epsilon,\epsilon}_{\rm EMf}(\pmb{P}_{2m'}||\pmb{H}_{n-2m'})=\sum_{\sigma}\,C^{\W\epsilon}_1(\sigma){\cal A}^\epsilon_{\rm sYMS}(\pmb{S}_{2m'}||\pmb{G}_{n-2m'};1,\sigma_2,\cdots,\sigma_{n-1},n)\nn
& &{\cal A}^{\W\epsilon}_{\rm EM}(\pmb{P}_{n}||\emptyset)=\sum_{\sigma}\,C^{\W\epsilon}_1(\sigma){\cal A}_{\phi^4}(1,\sigma_2,\cdots,\sigma_{n-1},n)\,.
\eea

Finally, we apply operators ${\cal T}^\epsilon[\pmb{\bar{1'}}]\cdots{\cal T}^\epsilon[\pmb{\bar{r'}}]$ and ${\cal T}^\epsilon_{{\cal X}_{2m'}}$ on two sides of the expansion of BI amplitudes in \eref{expansion for 12345} to obtain
\bea
& &{\cal A}^{\epsilon}_{\rm exDBI}(\pmb{\bar{1'}}|\cdots|\pmb{\bar{r'}}||\pmb{P}_{n-(|\pmb{1'}|+\cdots+|\pmb{r'}|)})=\sum_{\sigma}\,C_4(\sigma){\cal A}^\epsilon_{\rm YMS}(1,\sigma_2,\cdots,\sigma_{n-1},n)\,,\nn
& &{\cal A}^{\epsilon}_{\rm DBI}(\pmb{S}_{2m'}||\pmb{P}_{n-2m'})=\sum_{\sigma}\,C_4(\sigma){\cal A}^\epsilon_{\rm sYMS}(1,\sigma_2,\cdots,\sigma_{n-1},n)\,.~~~~\label{ex-12345-KK4}
\eea

Expansions obtained in this subsection, together with expansions for amplitudes of GR, EYM, EM and BI in \eref{expansion for 12345}, can be organized into the unified web as shown in Fig. \ref{web}. Notice that the general KK relations among amplitudes have not been included in the web. In this web, directions of arrows point to basis in the expansions, and different kinds of lines correspond to different kinds of coefficients: $C^\epsilon_1(\sigma)$, $C^\epsilon_2(\sigma)$, $C^\epsilon_{3a}(\sigma,\delta_{I_iI_j})$, $C^\epsilon_{3b}(\sigma)$ and $C_4(\sigma)$. This web emergences naturally from Table \ref{tab:unifying}, and includes same theories as the web for differential operators in \cite{Cheung:2017ems}.

\subsection{Double copy formula}

The unified web for expansions shows that amplitudes of all theories can be expanded to double color-ordered BAS amplitudes. To observe this, one can start with an arbitrary theory in the web in Fig. \ref{web}, and go through any line along the direction of arrow, then find that the end point is always the BAS theory.
This character can be manifested by organizing all expansions in the web into a compact double copy formula.

To arrive the double copy formula, we replace $\W\epsilon$ in the expansion of color-ordered YM amplitudes in \eref{ex-12345-KK2} by $\epsilon$, and substitute it into the expansion of GR amplitudes in \eref{expansion for 12345} to obtain
\bea
{\cal A}^{\epsilon,\W\epsilon}_{\rm GR}&=&\sum_{\sigma}\sum_{{\sigma}'}\,C^{\epsilon}_1(\sigma)C^{\W\epsilon}_1({\sigma}'){\cal A}_{\rm BAS}(1,\sigma_2,\sigma_3,\cdots,\sigma_{n-1},n;1,{\sigma}'_1,{\sigma}'_2,\cdots,{\sigma}'_{n-1},n)\,.~~~~\label{expand-GR-to-scalar}
\eea
In the above expression, GR amplitudes are expanded to BAS amplitudes which contain only propagators. Applying differential operators ${\cal O}^\epsilon$ and ${\cal O}^{\W\epsilon}$ on two sides of \eref{expand-GR-to-scalar} simultaneously, we get
\bea
{\cal O}^\epsilon{\cal O}^{\W\epsilon}{\cal A}^{\epsilon,\W\epsilon}_{\rm GR}&=&\sum_{\sigma}\sum_{{\sigma}'}\,\Big({\cal O}^\epsilon C^{\epsilon}_1(\sigma)\Big)\Big({\cal O}^{\W\epsilon} C^{\W\epsilon}_1({\sigma}')\Big){\cal A}_{\rm BAS}(1,\sigma_2,\sigma_3,\cdots,\sigma_{n-1},n;1,{\sigma}'_1,{\sigma}'_2,\cdots,{\sigma}'_{n-1},n)\,.
\eea
Thus, using relations provided in Table \ref{tab:unifying} and \eref{action of operator}, one can find the expansions of amplitudes to BAS amplitudes for all theories in the web in Fig. \ref{web}, formally unified as the double copy formula
\bea
{\cal A}=\sum_{\sigma}\sum_{{\sigma}'}\,{\cal C}(\sigma){\cal A}_{\rm BAS}(1,\sigma,n;1,\sigma',n){\cal C}(\sigma')\,,~~~~\label{ex-to-BS}
\eea
where numerators are given as ${\cal C}(\sigma)={\cal O}^\epsilon C^{\epsilon}_1(\sigma)$ and ${\cal C}(\sigma')={\cal O}^{\W\epsilon} C^{\W\epsilon}_1({\sigma}')$, while the propagator matrix is provided by double color-ordered BAS amplitudes.
For instance, since
\bea
{\cal A}_{\rm NLSM}(i'_1,\cdots,i'_n)=\big({\cal T}^\epsilon[i'_1\cdots i'_n]\big)\big({\cal L}^{\W\epsilon}\cdot{\cal T}^{\W\epsilon}[ab]\big){\cal A}^{\epsilon,\W\epsilon}_{\rm GR}(\pmb{H}_n)\,,
\eea
as shown in Table \ref{tab:unifying}, we get
\bea
{\cal A}_{\rm NLSM}(i'_1,\cdots,i'_n)&=&\sum_{\sigma}\sum_{{\sigma}'}\,\Big({\cal T}^\epsilon[i'_1\cdots i'_n] C^{\epsilon}_1(\sigma)\Big)\Big({\cal L}^{\W\epsilon}\cdot{\cal T}^{\W\epsilon}[ab] C^{\W\epsilon}_1({\sigma}')\Big){\cal A}_{\rm BAS}(1,\sigma,n;1,\sigma',n)\nn
&=&\sum_{\sigma}\sum_{{\sigma}'}\,C_5(\sigma){\cal A}_{\rm BAS}(1,\sigma,n;1,{\sigma}'_1,\sigma',n)C_4({\sigma}')\,,
\eea
where the relations in \eref{action of operator} were used in the second line.
Coefficients ${\cal C}(\sigma)$ and ${\cal C}(\sigma')$ for different theories are shown in Table \ref{tab:unifying2}.
\begin{table}[!h]
\begin{center}
\begin{tabular}{c|c|c}
Amplitude& ${\cal C}(\sigma)$  & ${\cal C}(\sigma')$ \\
\hline
${\cal A}_{{\rm GR}}^{\epsilon,\W\epsilon}(\pmb{H}_n)$ & $C_1^\epsilon(\sigma)$ & $C_1^{\W\epsilon}(\sigma')$  \\
${\cal A}_{{\rm EYM}}^{\epsilon,\W\epsilon}(\pmb{\bar{1}}|\cdots|\pmb{\bar{r}}||\pmb{H}_{n-(|\pmb{1}|+\cdots+|\pmb{r}|)})$ & $C_1^\epsilon(\sigma)$ & $C_2^{\W\epsilon}(\sigma')$ \\
${\cal A}_{{\rm EMf}}^{\epsilon,\W\epsilon}(\pmb{P}_{2m}||\pmb{H}_{n-2m})$& $C_1^\epsilon(\sigma)$ & $C_{3a}^{\W\epsilon}(\sigma',\delta_{I_iI_j})$  \\
${\cal A}_{{\rm EM}}^{\epsilon,\W\epsilon}(\pmb{P}_{2m}||\pmb{H}_{n-2m})$& $C_1^\epsilon(\sigma)$ & $C_{3b}^{\W\epsilon}(\sigma')$  \\
${\cal A}_{{\rm BI}}^\epsilon(\pmb{P}_n)$ & $C_1^\epsilon(\sigma)$ & $C_4(\sigma')$  \\
${\cal A}_{{\rm YM}}^\epsilon(i_1,\cdots,i_n)$& $C_1^\epsilon(\sigma)$ & $C_5(\sigma')$  \\
${\cal A}_{{\rm YMS}}^{\W\epsilon}(\pmb{\bar{1}}|\cdots|\pmb{\bar{r}}||\pmb{G}_{n-(|\pmb{1}|+\cdots+|\pmb{r}|)};i'_1,\cdots,i'_n)$ & $C_5(\sigma)$ & $C^{\W\epsilon}_2(\sigma')$\\
${\cal A}_{{\rm sYMS}}^{\W\epsilon}(\pmb{S}_{2m}||\pmb{G}_{n-2m};i'_1,\cdots,i'_n)$ & $C_5(\sigma)$ & $C^{\W\epsilon}_{3a}(\sigma',\delta_{I_iI_j})$ \\
${\cal A}_{\phi^4}(i'_1,\cdots,i'_n)$ & $C_5(\sigma)$ & $C^{\W\epsilon}_{3b}(\sigma')$\\
${\cal A}_{{\rm NLSM}}(i'_1,\cdots,i'_n)$ & $C_5(\sigma)$ & $C_4(\sigma')$ \\
${\cal A}_{{\rm BAS}}(i_1,\cdots,i_n;i'_1,\cdots,i'_n)$ & $C_5(\sigma)$ & $C_5(\sigma')$ \\
${\cal A}_{{\rm exDBI}}^{\W\epsilon}(\pmb{\bar{1}}|\cdots|\pmb{\bar{r}}||\pmb{P}_{n-(|\pmb{1}|+\cdots+|\pmb{r}|)})$ & $C_4(\sigma)$ & $C^{\W\epsilon}_2(\sigma')$ \\
${\cal A}_{{\rm DBI}}^{\W\epsilon}(\pmb{S}_{2m}||\pmb{P}_{n-2m})$ & $C_4(\sigma)$ & $C^{\W\epsilon}_{3a}(\sigma',\delta_{I_iI_j})$ \\
${\cal A}_{{\rm SG}}(\pmb{S}_n)$ & $C_4(\sigma)$ & $C_4 (\sigma')$ \\
\end{tabular}
\end{center}
\caption{\label{tab:unifying2}Unifying relations for expansions}
\end{table}

The double copy formula \eref{ex-to-BS} serves as the underlying foundation of the unified web for expansions in Fig. \ref{web}. All expansions in the unified web can be obtained from the double copy formula \eref{ex-to-BS}, by summing over $\sigma$ or $\sigma'$.
As an example, we consider the expansion of NLSM amplitudes
\bea
{\cal A}_{\rm NLSM}(\W\sigma_1,\W\sigma_2,\cdots,\W\sigma_n)=\sum_{\sigma}\sum_{{\sigma}'}\,{C}_5(\sigma){\cal A}_{\rm BAS}(1,\sigma,n;1,\sigma',n){ C}_4(\sigma')\,.~~~~\label{ex-NLSM1}
\eea
If we chose $\W\sigma_1=1$ and $\W\sigma_n=n$, we have ${C}_5(\sigma)=\delta_{\sigma\W\sigma}$, thus summing over $\sigma$ provides the expansion
\bea
{\cal A}_{\rm NLSM}(1,\W\sigma_2,\cdots,\W\sigma_{n-1},n)=\sum_{{\sigma}'}\,{\cal A}_{\rm BAS}(1,\W\sigma,n;1,\sigma',n){ C}_4(\sigma')\,,~~~~\label{ex-NLSM2}
\eea
which is the same as in \eref{ex-12345-KK2}. Substituting \eref{ex-NLSM2} into \eref{ex-NLSM1}, one find that summing over $\sigma'$ gives the generalized KK relations for color-ordered NLSM amplitudes which is given in \eref{ex-12345-KK3}. A more non-trivial example is the expansion of DBI amplitudes
\bea
{\cal A}_{\rm DBI}^{\W\epsilon}(\pmb{S}_{2m}||\pmb{P}_{n-2m})=\sum_{\sigma}\sum_{{\sigma}'}\,{C}_4(\sigma){\cal A}_{\rm BAS}(1,\sigma,n;1,\sigma',n){ C}_{3a}(\sigma',\delta_{I_iI_j})\,.~~~~\label{ex-exDBI1}
\eea
Similar with the derivation of \eref{ex-NLSM2}, one can summing over $\sigma$ in the formula
\bea
{\cal A}^{\W\epsilon}_{\rm sYMS}(\pmb{S}_{2m}||\pmb{G}_{n-2m};1,\W\sigma_2,\cdots,\W\sigma_{n-1},n)=\sum_{\sigma}\sum_{{\sigma}'}\,{C}_5(\sigma){\cal A}_{\rm BAS}(1,\sigma,n;1,\sigma',n){ C}_3^{\W\epsilon}(\sigma',\delta_{I_iI_j})
\eea
to obtain
\bea
{\cal A}^{\W\epsilon}_{\rm sYMS}(\pmb{S}_{2m}||\pmb{G}_{n-2m};1,\W\sigma_2,\cdots,\W\sigma_{n-1},n)=\sum_{{\sigma}'}\,{\cal A}_{\rm BAS}(1,\W\sigma,n;1,\sigma',n){ C}_3^{\W\epsilon}(\sigma',\delta_{I_iI_j})\,.~~~~\label{ex-exDBI2}
\eea
Substituting \eref{ex-NLSM2} into \eref{ex-exDBI1}, we find that summing over $\sigma$ gives
\bea
{\cal A}_{\rm DBI}^{\W\epsilon}(\pmb{S}_{2m}||\pmb{P}_{n-2m})=\sum_{{\sigma}'}\,{\cal A}_{\rm NLSM}(1,\sigma',n){ C}_{3a}(\sigma',\delta_{I_iI_j})\,,
\eea
which is given in \eref{ex-12345-KK3}. On the other hand, using \eref{ex-exDBI2} we get that summing over $\sigma'$ provides
\bea
{\cal A}_{\rm DBI}^{\W\epsilon}(\pmb{S}_{2m}||\pmb{P}_{n-2m})=\sum_{\sigma}\,{C}_4(\sigma){\cal A}^{\W\epsilon}_{\rm sYMS}(\pmb{S}_{2m}||\pmb{G}_{n-2m};1,\sigma,n)\,,
\eea
which is the last line in \eref{ex-12345-KK4}.

Table \ref{tab:unifying2} can be regarded as the dual version of Table \ref{tab:unifying} by following reasons. First at all, Table \ref{tab:unifying2} is derived only through information in Table \ref{tab:unifying}, without any other prior assumption. Secondly, they include same theories. Thirdly, operators ${\cal O}^\epsilon$ and coefficients ${\cal C}(\sigma)$ are linked by acting operators on BCJ numerators $C_1^\epsilon(\sigma)$ as ${\cal C}(\sigma)={\cal O}^\epsilon C_1^\epsilon(\sigma)$, and so do operators ${\cal O}^{\W\epsilon}$ and coefficients ${\cal C}(\sigma')$. Thus we have one to one dual mappings between operators and coefficients, as listed in Table \ref{tab:dual mapping}.
\begin{table}[!h]
\begin{center}
\begin{tabular}{c|c}
${\cal O}^\epsilon$  & ${\cal C}(\sigma)$ \\
\hline
$\mathbb{I}$ & $C_1^\epsilon(\sigma)$ \\
${\cal T}^\epsilon[\pmb{\bar{1}}]\cdots{\cal T}^\epsilon[\pmb{\bar{r}}]$ & $C_2^\epsilon(\sigma)$ \\
${\cal T}^\epsilon_{{\cal X}_{2m}}$ & $C_{3a}^\epsilon(\sigma,\delta_{I_iI_j})$ \\
${\cal T}^\epsilon_{ X_{2m}}$ & $C_{3b}^\epsilon(\sigma)$ \\
${\cal L}^\epsilon\cdot{\cal T}^\epsilon[ab]$ & $C_4(\sigma)$ \\
${\cal T}^\epsilon[i_1,\cdots,i_n]$ & $C_5(\sigma)$ \\
\end{tabular}
\end{center}
\caption{\label{tab:dual mapping}Dual mappings}
\end{table}

Since the duality between Table \ref{tab:unifying} and Table \ref{tab:unifying2} mentioned above, one can also conclude that the double copy formula \eref{ex-to-BS} is the dual version of the formula \eref{fund-uni-diff}, and the unified web for expansions in Fig. \ref{web} is the dual web of the unified web for differential operators in \cite{Cheung:2017ems}.

\section{Summary and discussion}
\label{summary}

In this paper, by deriving expansions of amplitudes for various theories via differential operators, and using results in literatures, we have constructed the unified web for expansions of amplitudes for a wide range of theories include GR, EYM, EM, BI, YM, YMS, $\phi^4$, NLSM, BAS, DBI and SG. Through out the whole work, we only use knowledge of the unified web for differential operators, without any other prior assumption. The new web for expansions is the dual version of the unified web for differential operators: two webs include same theories, and there exist one to one dual mappings between coefficients of basis and differential operators. Connections between amplitudes of various theories, which are indicated by CHY integrands and differential operators, are reflected by expansions in the current framework. All amplitudes in the new web can be expanded to double color-ordered BAS amplitudes. These expansions are unified into the double copy formula, with the propagator matrix formed by BAS amplitudes.

In \cite{Zhou:2018wvn}, a remaining question is that what theory does combinations of operators beyond Table \ref{tab:unifying} describe? This puzzle may be solved by the result in the current paper. Using the language of dual relations in Table \ref{tab:unifying2}, this question is transmuted to: which theories correspond to combinations of coefficients beyond Table \ref{tab:unifying2}, for example ${\cal C}(\sigma)=C^\epsilon_2(\sigma)$, ${\cal C}(\sigma')=C^{\W\epsilon}_{3b}(\sigma')$? The solution of this question seems visitable, since the explicit expression of the corresponding amplitudes can be calculated by the double copy formula \eref{ex-to-BS}. More explicitly, one can compute BAS amplitudes via Feynman rules or integration rules for single poles in the CHY framework \cite{Baadsgaard:2015voa,Baadsgaard:2015ifa}, and evaluate coefficients by applying algorithms in section \ref{coefficients}.
The solution also leads to CHY integrands for theories beyond in \cite{Cachazo:2014xea}, since one can derive these integrands by acting corresponding operators on the CHY integrand for GR.
Seeking the answer of this question is an interesting future direction.

\section*{Acknowledgments}

The author would thank Prof. Bo Feng for helpful discussions and valuable comments on the original manuscript. This
work is supported by Chinese NSF funding under
contracts No.11805163, as well as NSF of Jiangsu Province under Grant No.BK20180897.


\end{document}